\documentclass[11pt,draft]{article}

\usepackage[T1]{fontenc}

\usepackage{amsmath,amssymb}
\usepackage{amsthm}

\newcommand{\LL}{{\cal L}}

\renewcommand{\S}{{\cal S}}

\newcommand{\eps}{\epsilon}
\newcommand{\veps}{\varepsilon}

\newcommand{\ot}{\otimes}

\newcommand{\we}{\wedge}

\newcommand{\vth}{\vec{\theta}}
\newcommand{\lr}{\lrcorner}

\newcommand{\Star}[1]{*'\!\!_{#1}\,}

\newcounter{mnotecount}[section]

\numberwithin{equation}{section}
\numberwithin{thr}{section}

\begin{document}

\title{ADM-like Hamiltonian formulation of gravity in the teleparallel geometry: derivation of constraint algebra}
\author{ Andrzej Oko{\l}\'ow}
\date{September 18, 2013}

\maketitle
\begin{center}
{\it  Institute of Theoretical Physics, Warsaw University\\ ul. Ho\.{z}a 69, 00-681 Warsaw, Poland\smallskip\\
oko@fuw.edu.pl}
\end{center}
\medskip

\begin{abstract}
We derive a new constraint algebra for a Hamiltonian formulation of the Teleparallel Equivalent of General Relativity treated as a theory of cotetrad fields on a spacetime. The algebra turns out to be closed.
\end{abstract}

%***************************************************
\section{Introduction}
%***************************************************

In our previous paper \cite{oko-tegr-I} we presented a Hamiltonian formulation of the Teleparallel Equivalent of General Relativity (TEGR) regarded as a theory of cotetrad fields on a spacetime---the formulation is meant to serve as a point of departure for canonical quantization {\it \`a la Dirac} of the theory (preliminary stages of the quantization are described in \cite{q-suit,ham-nv,q-tegr}). In \cite{oko-tegr-I} we found a phase space, a set of (primary and secondary) constraints on the phase space and a Hamiltonian. We also presented an algebra of the constraints. An important fact is that this algebra is closed i.e. a Poisson bracket of every pair of the constraints is a sum of all the constraints multiplied by some factors. This property of the constraint algebra together with a fact  that the Hamiltonian is a sum of the constraints allowed us to conclude that $(i)$ the set of constraints is complete and $(ii)$ all the constraints are of the first class. 

Let us emphasize that knowledge of a complete set of constraints and their properties {as well as knowledge of an explicite form of a constraint algebra is very important from the point of view the Dirac's approach to canonical quantization of constrained systems since the knowledge enables a right treatment of constraints in the procedure of quantization.}  

However, the derivation of the constraint algebra turned out to be too long to be included in \cite{oko-tegr-I}. To fill this gap{, that is, to prove that the constraint algebra is correct} we carry out the derivation in the present paper\footnote{The derivation is also an example of an application of differential form calculus to a derivation of a constraint algebra which usually is done by means of tensor calculus.}. Moreover, to the best of our knowledge a derivation of constraint algebra of TEGR treated as a theory of cotetrad fields has never been presented before---in papers \cite{maluf-1,maluf} describing a distinct Hamiltonian formulation of this version\footnote{There is another version of TEGR configuration variables of which are cotetrad fields and flat Lorentz connections of non-zero torsion. For a complete Hamiltonian analysis of this version of TEGR see \cite{bl}.} of TEGR one can find a constraint algebra but its derivation is not shown. 

The paper is organized as follows: in Section 2 we recall the description of the phase space and the constraints on it derived in \cite{oko-tegr-I}. In Section 3 we derive the constraint algebra. Section 4 contains a short summary. 

{Let us finally emphasize that since the present paper plays a role of an appendix to \cite{oko-tegr-I} we do not discuss here the results nor compare them to results of previous works---all these can be found in \cite{oko-tegr-I}.}

%***************************************************
\section{Preliminaries}
%***************************************************

Let $\mathbb{M}$ be a four-dimensional oriented vector space equipped with a scalar product $\eta$ of signature $(-,+,+,+)$. We fix an orthonormal basis $(v_A)$ $(A=0,1,2,3)$ such that the components $(\eta_{AB})$ of $\eta$ given by the basis form a matrix ${\rm diag}(-1,1,1,1)$. The matrix $(\eta_{AB})$ and its inverse $(\eta^{AB})$ will be used to, respectively, lower and raise capital Latin letter indeces.     

Let $\Sigma$ be a three-dimensional oriented manifold. We assume moreover that it is a compact manifold without boundary. 

In \cite{oko-tegr-I} we obtained the phase space of TEGR as a Cartesian product of
\begin{enumerate}
\item a set of all quadruplets of one-forms $(\theta^{A})$ ($A=0,1,2,3$)  on $\Sigma$ such that for each quadruplet a metric
\begin{equation}
q:=\eta_{AB}\theta^A\ot\theta^B
\label{q-tt}
\end{equation}
on $\Sigma$ is {\em Riemannian} (i.e. positive definite); 
\item a set of all quadruplets $(p_B)$ ($B=0,1,2,3$)  of two-forms on $\Sigma$---a two-form $p_A$ is the momentum conjugate to $\theta^A$.
\end{enumerate}         
 
The metric $q$ defines a volume form $\eps$ on $\Sigma$ and a Hodge operator $*$ acting on differential forms on the manifold. Throughout the paper we will often use functions on $\Sigma$ defined as follows \cite{os}: 
\begin{equation}
\xi^A:=-\frac{1}{3!}\veps^A{}_{BCD}*(\theta^B\we\theta^C\we\theta^D),
\label{xi-sol}
\end{equation}
where $\veps_{ABCD}$ are components of a volume form on $\mathbb{M}$ given by the scalar product $\eta$.

Components of $q$ in a local coordinate frame $(x^i)$ ($i=1,2,3$) on $\Sigma$ will be denoted by $q_{ij}$. Obviously
\begin{equation}
q_{ij}=\eta_{AB}\theta^A_i\theta^B_j,
\label{q}
\end{equation}      
where $\theta^A_i$ are components of $\theta^A$. The metric $q$ and its inverse $q^{-1}$,
\begin{equation}
q^{-1}:=q^{ij}\partial_i\ot\partial_j, \ \ \ q^{ij}q_{jk}=\delta^i{}_k,
\label{q-1}
\end{equation}
will be used to, respectively, lower and raise, indeces (here: lower case Latin letters) of components of tensor fields defined on $\Sigma$. In particular we will often map one-forms to vector fields on $\Sigma$---a vector field corresponding to a one form $\alpha$ will be denoted by $\vec{\alpha}$ i.e. if $\alpha=\alpha_i dx^i$ then
\[
\vec{\alpha}:=q^{ij}\alpha_i\partial_j.
\]    

Let us emphasize that all object defined by $q$ (as $\eps$, $*$, $\xi^A$ and $q^{-1}$) are functions of $(\theta^A)$ which means that they are functions on the phase space.

In \cite{oko-tegr-I} we found some constraints on the phase space of TEGR. Smeared versions of the constraints read
\begin{align}
B(a):=&\int_\Sigma a\we(\theta^A\we*d\theta_A+\xi^Ap_A),\label{B-sm}\\
R(b):=&\int_\Sigma b\we(\theta^A\we*p_A-\xi^Ad\theta_A),\label{R-sm}\\
S(M):=&\int_\Sigma M\Big(\frac{1}{2}(p_A\we\theta^B)\we*(p_B\we\theta^A)-\frac{1}{4}(p_A\we\theta^A)\we*(p_B\we\theta^B)-\xi^A\we{d}p_A+\nonumber\\+&\frac{1}{2}(d\theta_A\we\theta^B)\we{*}(d\theta_B\we\theta^A)-\frac{1}{4}(d\theta_A\we\theta^A)\we{*}(d\theta_B\we\theta^B)\Big),\label{S-sm}\\
V(\vec{M}):=&\int_\Sigma -{d}{\theta}^A\we(\vec{M}\lrcorner p_A)-(\vec{M}\lrcorner{\theta}^A)\we {d}p_A,\label{V-sm}
\end{align}     
where $a,b,M$ and $\vec{M}$ are smearing fields on $\Sigma$: $a$ and $b$ are one-forms, $M$ is a function and $\vec{M}$ a vector field on the manifold. The smearing field possess altogether ten degrees of freedom per point of $\Sigma$. In \cite{oko-tegr-I} we called $B(a)$ {\em boost} constraint and $R(b)$ {\em rotation} constraint. $S(M)$ is a {\em scalar} constraint and $V(\vec{M})$ a {\em vector} constraint of TEGR.

%***************************************************
\section{Derivation of the constraint algebra}
%***************************************************

In this section we will calculate Poisson brackets of all pairs of the constraints presented above and show that each Poisson bracket is a sum of the constraints smeared with some fields. 

Calculations needed to achieve the goal will be long, laborious and complicated. We assume that the reader is familiar with tensor calculus, differential form calculus (including the contraction $\vec{X}\lr\alpha$ of a vector field $\vec{X}$ with a differential form $\alpha$) and properties of a Hodge operator on three-dimensional manifold defined by a Riemannian metric.  

%***************************************************
\subsection{Preliminaries}
%***************************************************

%***************************************************
\subsubsection{Poisson bracket}
%***************************************************

If $F$ and $G$ are functionals on the phase space then their Poisson bracket \cite{mielke}
\[
\{F,G\}=\int_\Sigma\Big(\frac{\delta F}{\delta {\theta}^A}\we\frac{\delta G}{\delta p_A}-\frac{\delta G}{\delta {\theta}^A}\we\frac{\delta F}{\delta p_A}\Big),
\]          
where the functional derivatives with respect to $\theta^A$ and $p_A$ are defined as follows \cite{os}: $\delta F/\delta\theta^A$ is a differential two-form on $\Sigma$ and  $\delta F/\delta p_A$ is a differential one-form on $\Sigma$ such that   
\[
\delta F=\int_\Sigma\delta\theta^A\we\frac{\delta F}{\delta \theta^A}+\delta p_A\we\frac{\delta F}{\delta p_A}
\]
for every $\delta\theta^A$ and $\delta p_A$.  

Calculating functional derivatives of the smeared constraints would be straightforward if $(i)$ the Hodge operator $*$ did not depend on $\theta^A$ and $(ii)$ the constraints did not depend on $\xi^A$ being a complicated function of $\theta^A$. Thus explicite formulae describing these derivatives are needed.

Given $k$-forms $\alpha$ and $\beta$, denote  
\begin{equation}
\alpha \we\Star{A}\beta\equiv\vec{\theta}^B\lrcorner[\eta_{AB}\,\alpha\we{*}\beta-(\vec{\theta}_A\lrcorner\alpha)\we{*}(\vec{\theta}_B\lrcorner\beta)-(\vec{\theta}_B\lrcorner\alpha)\we{*}(\vec{\theta}_A\lrcorner\beta)].
\label{a*'b}
\end{equation}
If the forms $\alpha,\beta$ do not depend on the canonical variables then \cite{os}
\begin{equation}
\frac{\delta}{\delta{\theta}^A}\int_\Sigma\alpha\we*\beta=\alpha\we\Star{A}\beta.
\label{a*b-fder}
\end{equation}
An important property of every two-form  $\alpha \we\Star{A}\beta$ is that it vanishes once contracted with the function $\xi^A$ \cite{os}:
\begin{equation}
\xi^A(\alpha\we\Star{A}\beta)=0.
\label{xi-d*=0}
\end{equation}

Consider now a three-form $\kappa_A$ on $\Sigma$ which does not depend on $\theta^A$ and $p_A$  and a functional 
\[
F=\int_\Sigma \xi^D\kappa_D=\int_\Sigma(*\xi^D)\we *(\kappa_D)=\int_\Sigma-\frac{1}{3!}\veps^D{}_{BCA}\theta^B\we\theta^C\we\theta^A\we*\kappa_D. 
\]
Using \eqref{a*b-fder} we obtain
\begin{multline*}
\delta F=\int_\Sigma\Big(-\frac{1}{3!}\veps^D{}_{BCA}(\delta\theta^B\we\theta^C\we\theta^A+\theta^B\we\delta\theta^C\we\theta^A+\theta^B\we\theta^C\we\delta\theta^A)\we*\kappa_D +\\+\delta\theta^A\we\big((*\xi^D)\we \Star{A}\kappa_D\big)\,\Big)
\end{multline*}
and therefore
\begin{equation}
\frac{\delta F}{\delta \theta^A}=-\frac{1}{2}\veps^D{}_{BCA}\theta^B\we\theta^C\we*\kappa_D +(*\xi^D)\we \Star{A}\kappa_D
\label{xi/th}
\end{equation}

%***************************************************
\subsubsection{Auxiliary formulae}
%***************************************************

Auxiliary formulae presented below will be used throughout the calculations. Except them we will need many other formulae which will be derived in the subsequent subsections. 

The functions $(\xi^A)$ satisfy the following important conditions \cite{nester}: 
\begin{align*}
\xi^A\xi_A&=-1, & \xi^A\theta_A&=0.
\end{align*}
These two equations imply
\begin{align*}
\xi^Ad\xi_A&=0, & d\xi^A\we\theta_A+\xi^Ad\theta_A&=0.
\end{align*}
These formulae will be used very often and therefore it would be troublesome to refer to them each time. Therefore we kindly ask the reader to keep the formulae in mind since they will be used without any reference. 

For any one-form $\alpha$ and any $k$-form $\beta$ \cite{os}
\begin{equation}
*(*\beta\we\alpha)=\vec{\alpha}\lr\beta.
\label{a-b}
\end{equation}
Setting $\beta=*\gamma$ and taking into account that $**={\rm id}$ we obtain an identity 
\begin{equation}
*(\gamma\we\alpha)=\vec{\alpha}\lr(*\gamma)
\label{a-*g}
\end{equation}
valid for every $l$-form $\gamma$.  

It was shown in \cite{oko-tegr-I} that
\begin{align}
\vec{\theta}^B\lrcorner{\theta}^A&=\eta^{AB}+\xi^A\xi^B, \label{tt-eta}\\
\theta^A\we(\vth_A\lr\alpha)&=k\alpha \label{tta=ka}
\end{align}
where $\alpha$ is a $k$-form on $\Sigma$.

%***************************************************
\subsubsection{Tensor calculus \label{t-calc}}
%***************************************************

Although our original wish was to carry out all necessary calculations using differential form calculus only we were forced in some cases to use tensor calculus. Below we gathered some expressions which will be applied repeatedly in the sequel.

Let $\nabla$ denote a covariant derivative on $\Sigma$ defined by the Levi-Civita connection given by the metric $q$. Consequently, by virtue of \eqref{q}
 \begin{equation}
0=\nabla_{i}q_{jk}=\nabla_{i}(\theta_{Aj}\theta^A_k)=(\nabla_{i}\theta_{Aj})\theta^A_k+\theta_{Aj}(\nabla_{i}\theta^A_k)
\label{0-nab-q}
\end{equation}
and
\begin{equation}
(\nabla_a\theta^{Bb})\theta_{Bb}=0.
\label{nab-tt}
\end{equation}
Note also that   
\begin{equation}
\nabla_a\eps_{ijk}=0
\label{nab-e-0}
\end{equation}
because $\eps$ is defined by $q$.
 
For any one-forms $\alpha$ and $\beta$ 
\begin{equation}
\begin{aligned}
d\alpha&=\nabla_a\alpha_b\,dx^a\we dx^b, & *d\alpha&=(\nabla_a\alpha_b)\eps^{ab}{}_c\,dx^c,\\
d\alpha\we\beta&=(\nabla_a\alpha_b)\beta_c\,dx^a\we dx^b\we dx^c, &  *(d\alpha\we\beta)&=(\nabla_a\alpha_b)\beta_c\eps^{abc},
\end{aligned} 
\label{d-nabla}
\end{equation}
If $\alpha$ is a one-form, $\beta$ a two-form and $\gamma$ a three-form then 
\begin{equation}
\begin{aligned}
*d\!*\alpha&=q^{ab}\nabla_a\alpha_b=\nabla^a\alpha_a,\\
*d\!*\beta&=q^{ab}\nabla_a\beta_{cb}dx^c=\nabla^b\beta_{cb}dx^c,\\ 
*d\!*\gamma&=q^{ab}\nabla_a\gamma_{dcb}dx^d\ot dx^c=\nabla^b\gamma_{dcb}dx^d\ot dx^c=\frac{1}{2}\nabla^b\gamma_{dcb}dx^d\we dx^c,
\end{aligned}
\label{delta-b}
\end{equation}
We will also apply the following identities (for a proof see e.g. \cite{os}):
\begin{align}
\eps_{ibc}\eps^{abc}&=2\delta^a{}_i, & \eps_{ijc}\eps^{abc}&=2\delta^{[a}{}_i\delta^{b]}{}_j, & \eps_{ijk}\eps^{abc}&=3!\delta^{[a}{}_i\delta^b{}_j\delta^{c]}{}_k.
\label{eps-delta}
\end{align}
and a formula 
\begin{equation}
\alpha\we*\beta=\frac{1}{k!}\alpha_{a_1\ldots a_k}\beta^{a_1\ldots a_k}\eps,
\label{*-df}
\end{equation}
where $\alpha$ and $\beta$ are $k$-forms.

%***************************************************
\subsection{Poisson brackets of $B(a)$ and $R(b)$}
%***************************************************

In this subsection we will calculate Poisson brackets $\{B(a),B(a')\}$, $\{R(b),R(b')\}$ and $\{B(a),R(b)\}$. 

%***************************************************
\subsubsection{Auxiliary formulae}
%***************************************************

The following formulae will be used while calculating the brackets:
\begin{gather}
b\we\alpha\we*(b'\we\beta)-(b\leftrightarrow b')=*(b\we b')\we\alpha\we\beta,\label{bab'b}\\
(\alpha\we\Star{A}\beta)\we*(b\we\theta^A)=0, \label{abbt}\\
\theta_A\we*(\alpha\we\theta^A)=(3-k)*\alpha, \label{t*at}\\
\eps_{DBCA}\theta^B\we\theta^C\we*(b\we\theta^A)=0, \label{eps-bt}\\
-\frac{1}{2}\eps^D{}_{BCA}\theta^B\we\theta^C \xi^A=*\theta^D \label{ett-xi}.
\end{gather}
In \eqref{bab'b} $\alpha,\beta,b,b'$ are one-forms, in \eqref{abbt} $\alpha$ and $\beta$ are $k$-forms, and $b$ is a one-form, in \eqref{t*at} $\alpha$ is a $k$-form, and in \eqref{eps-bt} $b$ is a one-form.    

\begin{proof}[Proof of \eqref{bab'b}] 
\begin{multline}
b\we\alpha\we*(b'\we\beta)-(b\leftrightarrow b')=b\we**[\alpha\we*(b'\we\beta)]-(b\leftrightarrow b')=\\=-b\we*[(\vec{\alpha}\lr b')\beta-b'\vec{\alpha}\lr\beta]-(b\leftrightarrow b')=\vec{\alpha}\lr(b\we b')\we*\beta=-(b\we b')\vec{\alpha}\lr*\beta=\\=(b\we b')\we*(\alpha\we\beta)=*(b\we b')\we(\alpha\we\beta),
\end{multline}    
where in the second step we used \eqref{a-b}, in the third the fact that $b\we*b'$ is symmetric in $b$ and $b'$ and finally in the fifth step we applied \eqref{a-*g}.
\end{proof}
\begin{proof}[Proof of \eqref{abbt}]
To prove \eqref{abbt} note that the two-form $\alpha\we\Star{A}\beta$ given by \eqref{a*'b} is of the form $\vth^B\lr\gamma_{AB}$, where the three-form $\gamma_{AB}$ is symmetric in $A$ and $B$: $\gamma_{AB}=\gamma_{BA}$. Thus
\[
(\alpha\we\Star{A}\beta)\we*(b\we\theta^A)=(\vth^B\lr\gamma_{AB})\we*(b\we\theta^A)=\gamma_{AB}\we\vth^B\lr*(b\we\theta^A)=\gamma_{AB}\we*(b\we\theta^A\we\theta^B),
\]        
where in the last step we used \eqref{a-*g}. But $(b\we\theta^A\we\theta^B)$ is antisymmetric in $A$ and $B$, hence \eqref{abbt} follows.  
\end{proof}

\begin{proof}[Proof of \eqref{t*at}]
\begin{equation*}
\theta_A\we*(\alpha\we\theta^A)=\theta_A\we\vth^A\lr*\alpha=(3-k)*\alpha,
\end{equation*}
where in the first step we used \eqref{a-*g} and in the second one we applied \eqref{tta=ka}.
\end{proof}
\begin{proof}[Proof of \eqref{eps-bt}]
Let us transform the following expression by means of \eqref{a-*g}: 
\begin{multline}
\theta^B\we\theta^C\we*(b\we\theta^A)=\theta^B\we\theta^C\we\vth^A\lr*b=-\vth^A\lr(\theta^B\we\theta^C)\we*b=\\=-(\vth^A\lr\theta^B)\we\theta^C\we*b+\theta^B(\vth^A\lr\theta^C)\we*b.
\label{ttbt}
\end{multline}
Note that the first term at the r.h.s. of the equation above is symmetric in $A$ and $B$, while the second one---in $A$ and $C$. This means that both terms vanish once contracted with $\eps_{DBCA}$.
\end{proof}

The last formula \eqref{ett-xi} is proven in \cite{os}.

%***************************************************
\subsubsection{Poisson bracket $\{B(a),B(a')\}$ }
%***************************************************

Let
\begin{align}
B_1(a)&:=\int_\Sigma a\we \theta^A\we*d\theta_A,& B_2(a)&:=\int_\Sigma a\we \xi^Ap_A=\int_\Sigma(*\xi^A)\we*(a\we p_A).
\label{B1B2}
\end{align}
Taking into account \eqref{B-sm} we see that $B(a)=B_1(a)+B_2(a)$ and consequently 
\begin{equation}
\{B(a),B(a')\}=\{B_1(a),B_1(a')\}+\Big(\{B_1(a),B_2(a')\}-\{B_1(a'),B_2(a)\}\Big)+\{B_2(a),B_2(a')\}.
\label{BB}
\end{equation}

Corresponding variational derivatives read
\begin{equation}
\begin{aligned}
&\frac{\delta B_1(a)}{\delta \theta^A }=-a\we*d\theta_A+(a\we\theta^B)\we\Star{A}d\theta_B+d*(a\we\theta_A),\\
&\frac{\delta B_1(a)}{\delta p_A}=0,\\
&\frac{\delta B_2(a)}{\delta \theta^A}=-\frac{1}{2}\eps^D{}_{BCA}\theta^B\we\theta^C *(a\we p_D)+(*\xi^B)\we\Star{A}(a\we p_B),\\
&\frac{\delta B_2(a)}{\delta p_A}=a\xi^A.
\end{aligned}
\label{B-var}
\end{equation}

Obviously, $\{B_1(a),B_1(a')\}=0$. The next term in \eqref{BB}
\begin{multline*}
\{B_1(a),B_2(a')\}-\{B_1(a'),B_2(a)\}=\int_\Sigma-a\we*d\theta_A\we a'\xi^A+d*(a\we\theta_A)\we a'\xi^A-(a\leftrightarrow a')=\\=\int_\Sigma2*(a\we a')\xi^Ad\theta_A-\Big(a\we\theta_A\we*(a'\we d\xi^A)-(a\leftrightarrow a')\Big)=\\=\int_\Sigma2*(a\we a')\xi^Ad\theta_A-*(a\we a')\theta_A\we d\xi^A=\int_\Sigma*(a\we a')\xi^Ad\theta_A,
\end{multline*} 
where in the first step we used \eqref{xi-d*=0} and in the third one \eqref{bab'b}. The last term in \eqref{BB} due to \eqref{xi-d*=0} reads
\[
\{B_2(a),B_2(a')\}=\int_\Sigma-\frac{1}{2}\eps^D{}_{BCA}\theta^B\we\theta^C\we*(a\we p_D)\we a'\xi^A-(a\leftrightarrow a').
\]
By virtue of \eqref{ett-xi} and \eqref{a-b}
\begin{equation*}
-\frac{1}{2}\eps^D{}_{BCA}\theta^B\we\theta^C\we*(a\we p_D)\we a'\xi^A=*\theta_D\we a'*(a\we p^D)=a\we p_D\vth^D\lr a'.
\end{equation*}
Thus 
\begin{multline*}
\{B_2(a),B_2(a')\}=-\int_\Sigma \vth^D\lr(a\we a')p_D=\int_\Sigma a\we a'\we\vth^D\lr p_D=\int_\Sigma-a\we a'\we *(\theta^D\we*p_D)=\\=\int_\Sigma-*(a\we a')\theta^A\we*p_A, 
\end{multline*}
where in the third step we used \eqref{a-b}. 

We conclude that (see \eqref{R-sm})
\begin{equation}
\{B(a),B(a')\}=-\int_\Sigma*(a\we a')\we(\theta^A\we*p_A-\xi^Ad\theta_A)=-R(*(a\we a')).
\label{BB-fin}
\end{equation}

%***************************************************
\subsubsection{Poisson bracket $\{R(b),R(b')\}$}
%***************************************************

Let us define
\begin{align}
R_1(b)&:=\int_\Sigma b\we\theta^A\we*p_A,& R_2(b)&:=\int_\Sigma b\we\xi^Ad\theta_A.
\label{R1R2}
\end{align}
Then by virtue of \eqref{R-sm} $R(b)=R_1(b)-R_2(b)$ and  
\begin{multline}
\{R(b),R(b')\}=\{R_1(b),R_1(b')\}-\Big(\{R_1(b),R_2(b')\}-\{R_1(b'),R_2(b)\}\Big)+\\+\{R_2(b),R_2(b')\}.
\label{RR}
\end{multline}
Corresponding variational derivatives read
\begin{equation}
\begin{aligned}
&\frac{\delta R_1(b)}{\delta \theta^A }=-b\we *p_A+(b\we\theta^B)\we\Star{A}p_B\\
&\frac{\delta R_1(b)}{\delta p_A}=*(b\we\theta^A),\\
&\frac{\delta R_2(b)}{\delta \theta^A}=d(b\xi_A)+(b\we d\theta^B)\we\Star{A}(*\xi_B)-\frac{1}{2}\eps_{DBCA}\theta^B\we\theta^C*(b\we d\theta^D)\\
&\frac{\delta R_2(b)}{\delta p_A}=0.
\end{aligned}
\label{R-var}
\end{equation}
Due to \eqref{abbt} the first bracket at the r.h.s. of \eqref{RR} reduces to
\[
\{R_1(b),R_1(b')\}=\int_\Sigma-b\we *p_A\we *(b'\we\theta^A)-(b\leftrightarrow b')=\int_\Sigma *(b\we b')\we \theta^A\we*p_A,
\]
where in the last step we used \eqref{bab'b}. Similarly, by virtue of \eqref{abbt} the next two brackets in \eqref{RR} reduce to
\begin{multline*}
\{R_1(b),R_2(b')\}-\{R_1(b'),R_2(b)\}=\\=\int_\Sigma[-d(b'\xi_A)+\frac{1}{2}\eps_{DBCA}\theta^B\we\theta^C*(b'\we d\theta^D)]\we*(b\we\theta^A)-(b\leftrightarrow b')=\\=\int_\Sigma [ b'\we d\xi^A\we*(b\we\theta^A)+\frac{1}{2}\eps_{DBCA}\theta^B\we\theta^C\we*(b\we\theta^A)*(b'\we d\theta^D)]-(b\leftrightarrow b').
\end{multline*}
Note now that due to \eqref{eps-bt} the term containing $\eps_{DBCA}$ vanishes. Therefore 
\[
\{R_1(b),R_2(b')\}-\{R_1(b'),R_2(b)\}=\int_\Sigma*(b\we b')\we\xi^Ad\theta_A,
\] 
where we applied \eqref{bab'b}. Because $\{R_2(b),R_2(b')\}=0$ we finally obtain
\begin{equation}
\{R(b),R(b')\}=\int_\Sigma*(b\we b')\we(\theta^A\we*p_A-\xi^Ad\theta_A)=R(*(b\we b')).
\label{RR-fin}
\end{equation}

%***************************************************
\subsubsection{Poisson bracket $\{B(a),R(b)\}$}
%***************************************************

Due to \eqref{B-sm}, \eqref{R-sm}, \eqref{B1B2} and \eqref{R1R2}
\begin{multline}
\{B(a),R(b)\}=\{B_1(a),R_1(b)\}+\{B_2(a),R_1(b)\}-\{B_1(a),R_2(b)\}-\\-\{B_2(a),R_2(b)\}.
\label{BR}
\end{multline}
Using \eqref{abbt} we immediately obtain
\begin{multline*}
\{B_1(a),R_1(b)\}=\int_\Sigma-a\we*d\theta_A\we*(b\we\theta^A)+d*(a\we\theta_A)\we*(b\we\theta^A)=\\=\int_\Sigma*(a\we b)\we\theta^A\we*d\theta_A-b\we*d\theta_A\we*(a\we\theta^A)+d*(a\we\theta_A)\we*(b\we\theta^A),
\end{multline*}
where we transformed the first term by means of  \eqref{bab'b}. Three terms constituting the next bracket in \eqref{BR} vanish by virtue of \eqref{eps-bt}, \eqref{abbt} and \eqref{xi-d*=0} and  consequently
\begin{equation*}
\{B_2(a),R_1(b)\}=\int_\Sigma b\we*p_A\we a\xi^A=\int_\Sigma*(a\we b)\we\xi^Ap_A.
\end{equation*}
The bracket $\{B_1(a),R_2(b)\}$ is obviously zero and   
\begin{multline*}
-\{B_2(a),R_2(b)\}=\int_\Sigma[(d(b\xi_A))-\frac{1}{2}\eps_{DBCA}\theta^B\we\theta^C*(b\we d\theta^D)]\we a\xi^A =\\=\int_\Sigma-db\we a+*\theta_D\we a\we*(b\we d\theta^D)=\int_\Sigma-db\we a +*a\we\theta_D\we*(db\we\theta^D)-\theta_D\we*a\we*d(b\we\theta^D)=\\=\int_\Sigma -db\we a+*a\we*db+d*(\theta^D\we*a)\we b\we\theta_D=\int_\Sigma*d*(\theta_D\we*a)\we*(b\we\theta^D),
\end{multline*}
where in the first step we used \eqref{xi-d*=0}, in the second one we applied \eqref{ett-xi} and in the fourth step \eqref{t*at}. Gathering the partial results we obtain 
\begin{multline}
\{B(a),R(b)\}=\int_\Sigma*(a\we b)(\theta^A\we*d\theta_A+\xi^Ap_A)-b\we*d\theta_A\we*(a\we\theta^A)+\\+[d*(a\we\theta_A)+*d*(\theta_A\we*a)]\we*(b\we\theta^A)=B(*(a\we b))+\\+\int_\Sigma -b\we*d\theta_A\we*(a\we\theta^A)+[**d*(a\we\theta_A)+*d*(\theta_A\we*a)]\we*(b\we\theta^A).
\label{BR-0}
\end{multline}

Let us show now that the integrand in the last line of \eqref{BR-0} is zero. The first term in the integrand can be expressed as follows:
\begin{multline*}
-b\we*d\theta_A\we*(a\we\theta^A)=-[b\we*d*(*\theta_A)]\we*[a\we\theta^A]=-\frac{1}{2}[b\we*d*(*\theta_A)]_{lj}[a\we\theta^A]^{lj}\eps\\=b_l(\nabla^k\theta^i_{A})\eps_{ijk}(a^j\theta^{Al}-a^l\theta^{Aj})\eps=\delta^l{}_n\eps_{ijk}b^n(\nabla^k\theta^i_{A})(a^j\theta^{A}_l-a_l\theta^{Aj})\eps
\end{multline*}
---the second equality holds by virtue of \eqref{*-df} and the third one due to \eqref{delta-b} and \eqref{nab-e-0}. Using \eqref{eps-delta} we rewrite $\delta^l{}_n$ by means of ``epsilons'' and continue transformations
\begin{multline*}
-b\we*d\theta_A\we*(a\we\theta^A)=\frac{1}{2}\eps_{nab}\eps^{lab}\eps_{ijk}b^n(\nabla^k\theta^i_{A})(a^j\theta^{A}_l-a_l\theta^{Aj})\eps=\\=\eps_{nab}(\delta^{l}{}_i\delta^{a}{}_j\delta^{b}{}_k+\delta^{a}{}_i\delta^{b}{}_j\delta^{l}{}_k+\delta^{b}{}_i\delta^{l}{}_j\delta^{a}{}_k)b^n(\nabla^k\theta^i_{A})(a^j\theta^{A}_l-a_l\theta^{Aj})\eps=\\=-a_ib^n \eps_{njk}(\nabla^k\theta^{i}_A)\theta^{Aj}\eps+a^j b^n \eps_{nij}(\nabla^k\theta_A^{i})\theta^A_k\eps-a_k b^n \eps_{nij}(\nabla^k\theta_A^{i})\theta^{Aj}\eps=\\=a^j b^n \eps_{jni}(\nabla^k\theta_A^{i})\theta^A_k\eps+a_kb^n \eps_{nij}\theta^{Aj}(\nabla^i\theta^{k}_A-\nabla^k\theta_A^{i})\eps
\end{multline*}
---here in the second step we applied \eqref{eps-delta} to express $\eps^{lab}\eps_{ijk}$ by means of ``deltas''. 

On the other hand the second term in the integrand
\begin{multline*}
[**d*(a\we\theta_A)+*d*(\theta_A\we*a)]\we*(b\we\theta^A)=\\=[\nabla_j(a^i\theta^{Aj}-a^j\theta^{Ai})+\nabla^i(\theta^{Aj}a_j)]\eps_{ikl}b^k\theta^l_A\eps=\\=a^ib^k\eps_{ikl}\theta_A^l(\nabla^j\theta^A_j)\eps+a_jb^k\eps_{ikl}\theta^{Al}(\nabla^i\theta^j_A-\nabla^j\theta^{i}_A)\eps
\end{multline*}
---the second equality holds by virtue of \eqref{*-df} and \eqref{nab-e-0}. 

Thus the sum of the last two terms in \eqref{BR-0} is equal to
\[
a^ib^k\eps_{ikl}[\theta^A_j(\nabla^j\theta_A^l)+\theta_A^l(\nabla^j\theta^A_j)]\eps=0,
\]
where we used \eqref{0-nab-q}. Finally
\begin{equation}
\{B(a),R(b)\}=B(*(a\we b)).
\label{BR-fin}
\end{equation}

%***************************************************
\subsection{Poisson bracket of $S(M)$ and $S(M')$ }
%***************************************************

Following \cite{os} we split the constraint $S(M)$ given by \eqref{S-sm} into three functionals 
\begin{equation}
\begin{aligned}
&S_1(M):=\int_\Sigma M[\frac{1}{2}(p_A\we\theta^B)\we*(p_B\we\theta^A)-\frac{1}{4}(p_A\we\theta^A)\we*(p_B\we\theta^B)]\\
&S_2(M):=-\int_\Sigma M\xi^A{d}p_A,\\
&S_3(M):=\int_\Sigma M[ \frac{1}{2}(d\theta_A\we\theta^B)\we{*}(d\theta_B\we\theta^A)-\frac{1}{4}(d\theta_A\we\theta^A)\we{*}(d\theta_B\we\theta^B)].
\end{aligned}
\label{S_i}
\end{equation}
Then
\begin{multline}
\{S(M),S(M')\}=\{S_1(M),S_1({M'})\}+\{S_2(M),S_2({M'})\}+\{S_3(M),S_3({M'})\}+\\\Big(\{S_1({M}),S_2({M'})\}+\{S_2({M}),S_3({M'})\}+\{S_3({M}),S_1({M'})\}-(M\leftrightarrow M')\Big).
\label{SMSM}
\end{multline}

%***************************************************
\subsubsection{Poisson brackets $\{S_i(M),S_j(M')\}$ }
%***************************************************

Functional derivatives of the functionals read:
\begin{align}
\frac{\delta S_1(M)}{\delta {\theta}^A}=&M\Big(p_B*(p_A\we\theta^B)-\frac{1}{2}p_A*(p_B\we\theta^B)+\nonumber\\&+\frac{1}{2}(p_C\we\theta^B)\we\Star{A}(p_B\we\theta^C)-\frac{1}{4}(p_B\we\theta^B)\we\Star{A}(p_C\we\theta^C)\Big),\label{S1/t}\\
\frac{\delta S_1(M)}{\delta p_A}=&M\Big(\theta^B*(p_B\we\theta^A)-\frac{1}{2}\theta^A*(p_B\we\theta^B)\Big),\label{S1/p}\\
\frac{\delta S_2(M)}{\delta {\theta}^A}=&M\Big(\frac{1}{2}\eps^{D}{}_{BCA}{\theta}^{B}\we{\theta}^{C}*dp_{D}-(*\xi^B)\we\Star{A}dp_B\Big),\label{S2/t}\\
\frac{\delta S_2(M)}{\delta p_A}=&{d}(M\xi^A)=\xi^AdM+Md\xi^A,\label{S2/p}\\
\frac{\delta S_3(M)}{\delta {\theta}^A}=&d\Big(M\theta^B*(d\theta_B\we\theta_A)-\frac{M}{2}\theta_A*(d\theta_B\we\theta^B)\Big)+M\Big(d\theta_B*(d\theta_A\we\theta^B)-\nonumber\\&-\frac{1}{2}d\theta_A*(d\theta_B\we\theta^B)+\frac{1}{2}(d\theta_C\we\theta^B)\we\Star{A}(d\theta_B\we\theta^C)-\nonumber\\-&\frac{1}{4}(d\theta_B\we\theta^B)\we\Star{A}(d\theta_C\we\theta^C)\Big),\label{S3/t}\\
\frac{\delta S_3(M)}{\delta p_A}=&0.
\label{S3/p}
\end{align}

Let us begin the calculations with the bracket $\{S_1(M),S_1(M')\}$:
\begin{equation*}
\{S_1(M),S_1(M')\}=\int_\Sigma \Big(\frac{\delta S_1(M)}{\delta {\theta}^A}\we \frac{\delta S_1(M')}{\delta p_A}-(M\leftrightarrow M')\Big)=0
\end{equation*}
because the first term under the integral is symmetric in $M$ and $M'$. It was shown in \cite{os} that
\begin{equation}
\{S_2(M),S_2(M')\}=-\int_\Sigma (\vec{m}\lr\theta^A)\we dp_A,
\label{s2s2}
\end{equation}
where
\[
m:=MdM'-M'dM.
\]
Because $S_3(M)$ does not depend on the momenta
\[
\{S_3(M),S_3(M')\}=0.
\] 

Next,
\begin{multline}
\{S_1(M),S_2(M')\}-(M\leftrightarrow M')=\int_\Sigma M[p_B*(p_A\we\theta^B)-\frac{1}{2}p_A*(p_B\we\theta^B)]\we \xi^AdM'-\\-(M\leftrightarrow M')=\int_\Sigma m\we p_B*(\xi^Ap_A\we\theta^B)-\frac{1}{2}m\we\xi^Ap_A*(p_B\we\theta^B),
\label{s1s2}
\end{multline}
where in the first step we removed terms symmetric in $M$ and $M'$ and applied \eqref{xi-d*=0}.  
\begin{multline}
\{S_2(M),S_3(M')\}-(M\leftrightarrow M')=\int_\Sigma-M'[d\theta_B*(d\theta_A\we\theta^B)-\frac{1}{2}d\theta_A*(d\theta_B\we\theta^B)]\we \xi^AdM-\\-(M\leftrightarrow M')=\int_\Sigma m\we d\theta_B*(\xi^Ad\theta_A\we\theta^B)-\frac{1}{2}m\we\xi^Ad\theta_A*(d\theta_B\we\theta^B)
\label{s2s3}
\end{multline}
---here in the first step there vanished terms symmetric in $M$ and $M'$ and one being an exact three-form (recall that $\Sigma$ is a compact manifold without boundary); two terms vanished due to \eqref{xi-d*=0}.    
\begin{multline}
\{S_3(M),S_1(M')\}-(M\leftrightarrow M')=\int_\Sigma dM[\theta^B*(d\theta_B\we\theta_A)-\frac{1}{2}\theta_A*(d\theta_B\we\theta^B)]\we \\ \we M'[\theta^C*(p_C\we\theta^A)-\frac{1}{2}\theta^A*(p_C\we\theta^C)]-(M\leftrightarrow M')=\\=\int_\Sigma-m\we[\theta^B\we\theta^C*(d\theta_B\we\theta_A*(p_C\we\theta^A))-\frac{1}{2}\theta_A*(p_C\we\theta^A)\we\theta^C*(d\theta_B\we\theta^B)-\\-\frac{1}{2}\theta^B\we\theta^A*(d\theta_B\we\theta_A)*(p_C\we\theta^C)]=\int_\Sigma-m\we\theta^B\we\theta^A*(d\theta_B\we*p_A)-\\-\frac{1}{2}m\we\theta^A\we*p_A*(d\theta_B\we\theta^B)+\frac{1}{2}m\we\theta^B\we*d\theta_B*(p_A\we\theta^A),
\label{s3s1}
\end{multline}
where in the first step some terms disappeared by virtue of their symmetricity in $M$ and $M'$; moreover, we used \eqref{t*at} in the last step.

Thus we obtain an explicite expression for the r.h.s. of \eqref{SMSM}:
\begin{multline}
\{S(M),S(M')\}=\int_\Sigma -(\vec{m}\lr\theta^A)\we dp_A+m\we p_B*(\xi^Ap_A\we\theta^B)-\frac{1}{2}m\we\xi^Ap_A*(p_B\we\theta^B)+\\+m\we d\theta_B*(\xi^Ad\theta_A\we\theta^B)-\frac{1}{2}m\we\xi^Ad\theta_A*(d\theta_B\we\theta^B)-m\we\theta^B\we\theta^A*(d\theta_B\we*p_A)-\\-\frac{1}{2}m\we\theta^A\we*p_A*(d\theta_B\we\theta^B)+\frac{1}{2}m\we\theta^B\we*d\theta_B*(p_A\we\theta^A).
\label{SMSM-0}
\end{multline} 

%***************************************************
\subsubsection{Isolating constraints}
%***************************************************

Our goal now is to isolate constraints at the r.h.s. of \eqref{SMSM-0}, that is, to show that the r.h.s. of \eqref{SMSM-0} is a sum of the constraints \eqref{B-sm}--\eqref{V-sm} smeared with appropriately chosen fields.

It is clear (see \eqref{V-sm}) that the first term at the r.h.s. of \eqref{SMSM-0}
\[
\int_\Sigma -(\vec{m}\lr\theta^A)\we dp_A=V(\vec{m})+\int_\Sigma d\theta^A\we \vec{m}\lr p_A.
\]
The second and the third terms are equal to 
\begin{multline*}
\int_\Sigma[\theta^B*(m\we p_B)-\frac{1}{2}m*(p_B\we\theta^B)]\we\xi^A p_A=B\Big(\theta^B*(m\we p_B)-\frac{1}{2}m*(p_B\we\theta^B)\Big)+\\+\int_\Sigma -*(m\we p_B)\theta^B\we\theta^A\we* d\theta_A+\frac{1}{2}*(p_B\we\theta^B)m\we\theta^A\we*d\theta_A.
\end{multline*}
Similarly, the fourth and the fifth ones are equal to
\begin{multline*}
\int_\Sigma [\theta^B*(m\we d\theta_B)-\frac{1}{2}m*(d\theta_B\we\theta^B)]\we\xi^Ad\theta_A=-R\Big(\theta^B*(m\we d\theta_B)-\frac{1}{2}m*(d\theta_B\we\theta^B)\Big)+\\+\int_\Sigma*(m\we d\theta_B)\theta^B\we\theta^A\we*p_A-\frac{1}{2}*(d\theta_B\we\theta^B)m\we\theta^A\we*p_A. 
\end{multline*}
Thus \eqref{SMSM-0} can be rewritten as follows: 
\begin{multline}
\{S(M),S(M')\}=V(\vec{m})+B\Big(\theta^B*(m\we p_B)-\frac{1}{2}m*(p_B\we\theta^B)\Big)-\\-R\Big(\theta^B*(m\we d\theta_B)-\frac{1}{2}m*(d\theta_B\we\theta^B)\Big)+\text{remaining terms},
\label{SMSM-1}
\end{multline}
where the remaining terms read
\begin{multline}
\int_\Sigma -*(m\we p_B)\theta^B\we\theta^A\we* d\theta_A+*(m\we d\theta_B)\theta^B\we\theta^A\we*p_A+\\+*(p_B\we\theta^B)m\we\theta^A\we*d\theta_A-*(d\theta_B\we\theta^B)m\we\theta^A\we*p_A-\\-m\we\theta^B\we\theta^A*(d\theta_B\we*p_A)+d\theta^A\we \vec{m}\lr p_A.
\label{SMSM-rem}
\end{multline}

Now we will transform the remaining terms \eqref{SMSM-rem} to a form which will be a convenient starting point for isolating constraints. The first term in \eqref{SMSM-rem} 
\begin{multline}
-*(m\we p_B)\theta^B\we\theta^A\we* d\theta_A=-m\we p_B*(\theta^B\we\theta^A\we* d\theta_A)=\\=-m\we p_B\vth^B\lr*(\theta^A\we*d\theta_A)=-\vth^B\lr(m\we p_B)\we*(\theta^A\we*d\theta_A)=\\=-(\vth^B\lr m)p_B\we*(\theta^A\we*d\theta_A)+m\we *(*p_B\we\theta^B)\we*(\theta^A\we*d\theta_A)=\\=-(\vth^B\lr m)\theta^A\we*d\theta_A\we*p_B+m\we*(\theta^A\we*d\theta_A)\we*(\theta^B\we*p_B),
\label{SMSM-rem-t1}
\end{multline}
where we used \eqref{a-*g} in the second step and \eqref{a-b} in the fourth one. Similarly, the second term in \eqref{SMSM-rem}
\begin{multline}
*(m\we d\theta_B)\theta^B\we\theta^A\we* p_A=\\=(\vth^B\lr m)\we \theta^A\we*p_A\we*d\theta_B-m\we *(\theta^A\we*p_A)\we*(\theta^B\we*d\theta_B).
\label{SMSM-rem-t2}
\end{multline}
Next, we transform the third term in \eqref{SMSM-rem}:
\begin{multline}
m\we\theta^A\we*d\theta_A*(p_B\we\theta^B)=m\we\theta^A\we*d\theta_A\,\vth^B\lr*p_B =\vth^B\lr(m\we\theta^A\we*d\theta_A)\we*p_B=\\=(\vth^B\lr m)\theta^A\we*d\theta_A\we*p_B-m\we(\vth^B\lr\theta^A)*d\theta_A\we*p_B+\\+m\we\theta^A(\vth^B\lr*d\theta_A)\we*p_B=(\vth^B\lr m)\we\theta^A\we*d\theta_A\we*p_B-\\-m\we*d\theta^A\we*p_A-m\we*(\xi^Ad\theta_A)\we*(\xi^Bp_B)+m\we\theta^A\we*p_B*(d\theta_A\we\theta^B),
\label{SMSM-rem-t3}
\end{multline}
where in the first step we used \eqref{a-*g}, in the fourth one \eqref{tt-eta} and in the last one \eqref{a-*g} again. Similarly, the fourth term in \eqref{SMSM-rem}
\begin{multline}
-m\we\theta^A\we*p_A*(d\theta_B\we\theta^B)=-(\vth^B\lr m)\theta^A\we*p_A\we*d\theta_B+\\+m\we(\vth^B\lr\theta^A)*p_A\we*d\theta_B-m\we\theta^A\we*d\theta_B*(p_A\we\theta^B).
\label{SMSM-rem-t4}
\end{multline}
By virtue of \eqref{a-b} the last term in \eqref{SMSM-rem} can be expresses as follows \cite{os}:
\begin{equation}
d\theta^A\we \vec{m}\lr p_A=d\theta^A\we*(*p_A\we m)=*p_A\we m\we*d\theta^A=m\we *d\theta^A\we*p_A.
\label{SMSM-rem-tl}
\end{equation}
Now it is easy to see that the following pairs
\begin{enumerate}
\item the first term at the r.h.s. of \eqref{SMSM-rem-t1} and the first term at the r.h.s. of \eqref{SMSM-rem-t3},  
\item the first term at the r.h.s. of \eqref{SMSM-rem-t2} and the first term at the r.h.s. of \eqref{SMSM-rem-t4},
\item the second term at the r.h.s. of \eqref{SMSM-rem-t3} and \eqref{SMSM-rem-tl}
\end{enumerate}
sum up to zero. Moreover, the second term at the r.h.s. of  \eqref{SMSM-rem-t1} is equal to the second term at the r.h.s. of \eqref{SMSM-rem-t2}. Consequently, the terms \eqref{SMSM-rem} can be expresses as
\begin{multline}
\int_\Sigma (\vth^B\lr \theta^A)m\we*p_A\we*d\theta_B-m\we\theta^A\we*d\theta_B*(p_A\we\theta^B)+m\we\theta^A\we*p_B*(d\theta_A\we\theta^B)+\\+m\we*(\xi^Ap_A)\we*(\xi^Bd\theta_B)-m\we\theta^B\we\theta^A*(d\theta_B\we*p_A)+2m\we*(\theta^A\we*d\theta_A)\we*(\theta^B\we*p_B).
\label{SMSM-rem-3}
\end{multline}
Note now that the term above containing $\xi^A$ can be transformed as follows: 
\begin{multline}
\int_\Sigma m\we*(\xi^Ap_A)\we*(\xi^Bd\theta_B)=\int_\Sigma -*[m\we*(\xi^Bd\theta_B)]\we(\xi^Ap_A)=-B\Big(*(m\we\xi^B*d\theta_B)\Big)+\\+\int_\Sigma *[m\we*(\xi^Bd\theta_B)]\we \theta^A\we*d\theta_A=-B\Big(*(m\we\xi^B*d\theta_B)\Big) -\\-\int_\Sigma*[m\we*(\theta^A\we*d\theta_A)]\we\xi^Bd\theta_B=-B\Big(*(m\we\xi^B*d\theta_B)\Big)+R\Big(*[m\we*(\theta^A\we*d\theta_A)]\Big)-\\-\int_\Sigma m\we*(\theta^A\we*d\theta_A)\we*(\theta^B\we*p_B).
\label{xi-BR}
\end{multline}
Gathering the result above, \eqref{SMSM-rem-3} and \eqref{SMSM-1} we obtain 
\begin{multline}
\{S(M),S(M')\}=V(\vec{m})+B\Big(\theta^B*(m\we p_B)-\frac{1}{2}m*(p_B\we\theta^B)\Big)-\\-R\Big(\theta^B*(m\we d\theta_B)-\frac{1}{2}m*(d\theta_B\we\theta^B)\Big)-\\-B\Big(*(m\we\xi^B*d\theta_B)\Big)+R\Big(*[m\we*(\theta^A\we*d\theta_A)]\Big)+\text{remaining terms},
\label{SMSM-2}
\end{multline}
where the remaining terms read now
\begin{multline}
\int_\Sigma (\vth^B\lr \theta^A)m\we*p_A\we*d\theta_B-m\we\theta^A\we*d\theta_B*(p_A\we\theta^B)+m\we\theta^A\we*p_B*(d\theta_A\we\theta^B)-\\-m\we\theta^B\we\theta^A*(d\theta_B\we*p_A)+m\we*(\theta^A\we*d\theta_A)\we*(\theta^B\we*p_B).
\label{SMSM-rem-4}
\end{multline}

Now let us show that the remaining terms \eqref{SMSM-rem-4} can be expressed as $R(b)$ with the one-form $b$ being a complicated function of the canonical variables and $m$. By shifting the contraction $\vth^B\lr$ in the first term above and using \eqref{a-*g} one can easily show that the sum of the first and the second terms in \eqref{SMSM-rem-4} reads
\begin{multline}
\int_\Sigma(\vth^B\lr \theta^A)m\we*p_A\we*d\theta_B-m\we\theta^A\we*d\theta_B*(p_A\we\theta^B)=\\=\int_\Sigma\vth^B\lr (m\we*d\theta_B)\we\theta^A\we*p_A=R\Big(\vth^B\lr (m\we*d\theta_B)\Big)+\int_\Sigma \vth^B\lr (m\we*d\theta_B)\we\xi^Ad\theta_A.
\label{SMSM-rem-4-1,2}
\end{multline}

Let us now express the third term in \eqref{SMSM-rem-4} by means of the components of the canonical variables and the covariant  derivative $\nabla_a$ (see \eqref{d-nabla}):
\begin{multline*}
m\we\theta^A\we*p_B*(d\theta_A\we\theta^B)=m\we\frac{1}{2}\theta^A_ip_{Bjk}\eps^{jk}{}_ldx^i\we dx^l(\nabla_a\theta_{Ab})\theta^B_c\eps^{abc}=\\=m\we\theta^A_ip_{B}{}^{ab}dx^i\we dx^c(\nabla_a\theta_{Ab})\theta^B_c+m\we\theta^A_ip_{B}{}^{bc}dx^i\we dx^a(\nabla_a\theta_{Ab})\theta^B_c+\\+m\we\theta^A_ip_{B}{}^{ca}dx^i\we dx^b(\nabla_a\theta_{Ab})\theta^B_c,
\end{multline*}    
where in the last step we used \eqref{eps-delta}. Similarly, the fourth term in \eqref{SMSM-rem-4}
\[
-m\we\theta^B\we\theta^A*(d\theta_B\we*p_A)=-m\we\theta^B_i\theta^A_cdx^i\we dx^c(\nabla_a\theta_{Bb})p_{A}{}^{ab}
\] 
and consequently the sum of the third and the fourth terms in \eqref{SMSM-rem-4} is of the following form: 
\begin{multline*}
m\we\theta^A\we*p_B*(d\theta_A\we\theta^B)-m\we\theta^B\we\theta^A*(d\theta_B\we*p_A)=\\=m\we\theta^A_ip_{B}{}^{bc}dx^i\we dx^a(\nabla_a\theta_{Ab})\theta^B_c+m\we\theta^A_ip_{B}{}^{ca}dx^i\we dx^b(\nabla_a\theta_{Ab})\theta^B_c=\\=-m\we(\theta^{Bc}p_{Bc}{}^b)(d\theta_A)_{ba}dx^a\we\theta^A=-m\we[\overrightarrow{(\vth^B\lr p_B)}\lr d\theta^A]\we\theta^A=\\=-m\we*[*d\theta_A\we*(*p_B\we\theta^B)]\we\theta^A=-*[*(m\we\theta^A)\we*d\theta_A]\we\theta^B\we*p_B
\end{multline*}
---here in the fourth step we used \eqref{a-b}. Integrating the equation above over $\Sigma$ we obtain:
\begin{multline}
\int_\Sigma m\we\theta^A\we*p_B*(d\theta_A\we\theta^B)-m\we\theta^B\we\theta^A*(d\theta_B\we*p_A)=\\=R\Big(-*[*(m\we\theta^B)\we*d\theta_B]\Big)+\int_\Sigma-*[*(m\we\theta^B)\we*d\theta_B]\we \xi^Ad\theta_A.
\label{SMSM-rem-4-3,4}
\end{multline}
Finally, the last term in \eqref{SMSM-rem-4} 
\begin{multline}
\int_\Sigma m\we*(\theta^B\we*d\theta_B)\we*(\theta^A\we*p_A)=\int_\Sigma *[m\we*(\theta^B\we*d\theta_B)]\we \theta^A\we*p_A=\\=R\Big(*[m\we*(\theta^B\we*d\theta_B)]\Big)+\int_\Sigma *[m\we*(\theta^B\we*d\theta_B)]\we\xi^Ad\theta_A.
\label{SMSM-rem-4-5}
\end{multline}

Gathering the three results \eqref{SMSM-rem-4-1,2}, \eqref{SMSM-rem-4-3,4} and \eqref{SMSM-rem-4-5} we conclude that the terms \eqref{SMSM-rem-4} can be expressed as
\begin{multline}
R\Big(\vth^B\lr (m\we*d\theta_B)-*[*(m\we\theta^B)\we*d\theta_B]+*[m\we*(\theta^B\we*d\theta_B)]\Big)+\\+\text{terms independent of $p_A$},
\end{multline}
where the terms independent of $p_A$ read
\begin{multline}
\int_\Sigma \vth^B\lr (m\we*d\theta_B)\we\xi^Ad\theta_A-*[*(m\we\theta^B)\we*d\theta_B]\we \xi^Ad\theta_A+*[m\we*(\theta^B\we*d\theta_B)]\we\xi^Ad\theta_A.
\label{ind}
\end{multline} 

Now it is enough to show that the terms \eqref{ind} sums up to zero for all $m$ and $\theta^A$. To this end let us isolate the factor $m\xi^A$ in each term of \eqref{ind}---using \eqref{a-*g} and \eqref{a-b} we obtain 
\begin{multline*}
\int_\Sigma -m\xi^A\we*d\theta_B\we \vth^B\lr d\theta_A-(\vth^B\lr *m)\we*d\theta_B\we \xi^A*d\theta_A-m\we\vth^B\lr d\theta_B\we\xi^A*d\theta_A=\\=\int_\Sigma m\xi^A\we\big[-*d\theta_B\we \vth^B\lr d\theta_A+*(\vth^B\lr(*d\theta_B\we*d\theta_A))-\vth^B\lr d\theta_B\we*d\theta_A\big].
\end{multline*} 
Consider now the terms in the square bracket above:
\begin{multline*}
-*d\theta_B\we \vth^B\lr d\theta_A+*(\vth^B\lr(*d\theta_B\we*d\theta_A))-\vth^B\lr d\theta_B\we*d\theta_A=\\=-*d\theta_B\we \vth^B\lr d\theta_A+(\vth^B\lr *d\theta_B)d\theta_A-d\theta_B\vth^B\lr*d\theta_A-\vth^B\lr d\theta_B\we*d\theta_A=\\=\vth^B\lr(*d\theta_B\we d\theta_A)-\vth^B\lr(d\theta_B\we *d\theta_A)=\vth^B\lr(*d\theta_B\we d\theta_A)-\vth^B\lr(*d\theta_B\we d\theta_A)=0.
\end{multline*}
This means that indeed \eqref{ind} is zero for all $m$ and $\theta^A$. Consequently,
\begin{equation}
\text{the terms \eqref{SMSM-rem-4}}=R\Big(\vth^B\lr (m\we*d\theta_B)-*[*(m\we\theta^B)\we*d\theta_B]+*[m\we*(\theta^B\we*d\theta_B)]\Big).
\label{terms-R}
\end{equation}
Setting this result to \eqref{SMSM-2} we obtain 
\begin{multline}
\{S(M),S(M')\}=V(\vec{m})+\\+B\Big(\theta^B*(m\we p_B)-\frac{1}{2}m*(p_B\we\theta^B)\Big)-R\Big(\theta^B*(m\we d\theta_B)-\frac{1}{2}m*(d\theta_B\we\theta^B)\Big)-\\-B\Big(*(m\we\xi^B*d\theta_B)\Big)+R\Big(*[m\we*(\theta^A\we*d\theta_A)]\Big)+\\+R\Big(\vth^B\lr (m\we*d\theta_B)-*[*(m\we\theta^B)\we*d\theta_B]+*[m\we*(\theta^B\we*d\theta_B)]\Big)
\label{SMSM-alm-fin}
\end{multline}

%***************************************************
\subsubsection{Another form of $\{S(M),S(M')\}$}
%***************************************************

Let us now transform the result to a form in which both constraints $B$ given by \eqref{B-sm} and $R$ defined by \eqref{R-sm} appear on an equal footing. Consider the following transformation:
\begin{equation}
p_A\mapsto -d\theta_A, \ \ \ \ d\theta_A\mapsto p_A.
\label{p->dt}
\end{equation}
It is easy to see that under this transformation
\begin{align*}
&B(a(p_A,d\theta_B))\mapsto R(a(-d\theta_A,p_B)),\\
&R(b(p_A,d\theta_B))\mapsto -B(b(-d\theta_A,p_B))
\end{align*}

Let us now express the formula \eqref{xi-BR} in the following form:
\begin{multline*}
\int_\Sigma m\we*(\xi^Ap_A)\we*(\xi^Bd\theta_B)+m\we*(\theta^A\we*d\theta_A)\we*(\theta^B\we*p_B)=\\=-B\Big(*(m\we\xi^B*d\theta_B)\Big)+R\Big(*[m\we*(\theta^A\we*d\theta_A)]\Big).
\end{multline*}   
Note now that the l.h.s. of the identity above is invariant with respect to the transformation \eqref{p->dt}. Consequently, the r.h.s. has to be invariant too. Thus the fourth and the fifth terms in \eqref{SMSM-alm-fin}
\begin{multline}
-B\Big(*(m\we\xi^B*d\theta_B)\Big)+R\Big(*[m\we*(\theta^A\we*d\theta_A)]\Big)=-R\Big(*(m\we\xi^B*p_B)\Big)-\\-B\Big(*[m\we*(\theta^A\we*p_A)]\Big)=
-\frac{1}{2}B\Big(*(m\we\xi^B*d\theta_B)+*[m\we*(\theta^A\we*p_A)]\Big)+\\+\frac{1}{2}R\Big(-*(m\we\xi^B*p_B)+*[m\we*(\theta^A\we*d\theta_A)]\Big),
\label{-B+R}
\end{multline}
where the last equation holds by virtue of the following trivial fact: if $x=y$ then $x=\frac{1}{2}(x+y)$. Similarly, the expression \eqref{SMSM-rem-4} is also invariant with respect to \eqref{p->dt}. Thus by virtue of the identity \eqref{terms-R} the last term in \eqref{SMSM-alm-fin} 
\begin{multline}
R\Big(\vth^B\lr (m\we*d\theta_B)-*[*(m\we\theta^B)\we*d\theta_B]+*[m\we*(\theta^B\we*d\theta_B)]\Big)=\\=-B\Big(\vth^B\lr (m\we*p_B)-*[*(m\we\theta^B)\we*p_B]+*[m\we*(\theta^B\we*p_B)]\Big)=\\=\frac{1}{2}R\Big(\vth^B\lr (m\we*d\theta_B)-*[*(m\we\theta^B)\we*d\theta_B]+*[m\we*(\theta^B\we*d\theta_B)]\Big)-\\-\frac{1}{2}B\Big(\vth^B\lr (m\we*p_B)-*[*(m\we\theta^B)\we*p_B]+*[m\we*(\theta^B\we*p_B)]\Big).
\label{R(...)}
\end{multline}
Setting the results \eqref{-B+R} and \eqref{R(...)} to \eqref{SMSM-alm-fin} after some simple calculations with application of \eqref{a-b} and \eqref{a-*g} we obtain the final form of the Poisson bracket of the scalar constraints
\begin{multline}
\{S(M),S(M')\}=V(\vec{m})+B\Big(\theta^B*(m\we p_B)-\frac{1}{2}*(m\we\xi^B*d\theta_B)-\\-*[m\we*(\theta^B\we*p_B)]-\frac{1}{2}*(*m\we\theta^B)*p_B+\frac{1}{2}*[*(m\we\theta^B)\we*p_B]\Big)+\\+R\Big(-\theta^B*(m\we d\theta_B)-\frac{1}{2}*(m\we\xi^B*p_B)+\\+*[m\we*(\theta^B\we*d\theta_B)]+\frac{1}{2}*(*m\we\theta^B)*d\theta_B-\frac{1}{2}*[*(m\we\theta^B)\we*d\theta_B]\Big).
\label{SS-fin}
\end{multline}
Note that the sum of the constraints $B$ and $R$ at the r.h.s. of \eqref{SS-fin} is explicitely invariant with respect to the transformation \eqref{p->dt}.  

%***************************************************
\subsection{Poisson bracket of $R(b)$ and $S(M)$  }
%***************************************************

Recall that the constraints $R(b)$ and $S(M)$ are defined by, respectively, \eqref{R-sm} and \eqref{S-sm}. To show that the bracket $\{R(b),S(M)\}$ is a sum of the constraints \eqref{B-sm}--\eqref{V-sm} smeared with some fields we will proceed according to the following prescription. The bracket under consideration can be expressed as
\[
\{R(b),S(M)\}=\sum_{i=1}^2\sum_{j=1}^3 (-1)^{i+1}\{R_i(b),S_j(M)\},
\]
where the functionals at the r.h.s. are given by \eqref{R1R2} and \eqref{S_i}. It is not difficult to see that each bracket in the sum is either quadratic in $p_A$, linear in $p_A$ or independent of $p_A$. So we will first calculate the brackets and then we will gather similar terms according to the classification. Next, it will turn out that the terms quadratic in $p_A$ can be re-expressed as a constraint plus a term linear in $p_A$. Then it will turn out that all the linear terms can be re-expressed as some constraints plus a term independent of $p_A$. Finally we will show that all the term independent of $p_A$ sum up to zero.  

Except the prescription we will need some formulae and identities which will make easier the calculations.

%***************************************************
\subsubsection{Auxiliary formulae}  
%***************************************************

The following formulae will be used in the sequel while calculating both $\{R(b),S(M))\}$ and $\{B(a),S(M)\}$:
\begin{align}
\vth_A\lr*\xi^B&=-\frac{1}{2}\eps^B{}_{CDA}\theta^C\we\theta^D+\xi_A*\theta^B\label{vt-xi},\\
-\frac{1}{2}\eps^A{}_{BCD}\theta^B\we\theta^C\we\alpha&=-\xi_D\alpha\we*\theta^A+(*\xi^A)\vth_D\lr\alpha,\label{ett-al}\\
(\alpha\we\Star{A}\beta)\we\gamma&=\alpha\we*\beta\,\vth_A\lr\gamma-[(\vth_A\lr\alpha)\we*\beta+(\alpha\leftrightarrow\beta)]\we\gamma=\nonumber\\ &=[(-1)^k\alpha\we(\vth_A\lr*\beta)-(\vth_A\lr\beta)\we*\alpha]\we\gamma,\label{abg}\\
(\alpha\we\Star{A}\beta)\we\frac{\delta S_1(M)}{\delta p_A}&=M(k-\frac{5}{2})\alpha\we*\beta*(p_A\we\theta^A)+\nonumber\\&+(-1)^{4-k}M[*(\alpha\we\theta^A)\we*p_A\we\beta+(\alpha\leftrightarrow\beta)],\label{a*bs_1}\\
\alpha^A\we\beta_A&=\theta_A\we\vth^B\lr\alpha^A\we\beta_B+(-1)^k\theta_A\we\alpha^A\we\vth^B\lr\beta_B-\nonumber\\&-\xi_A\alpha^A\we\xi^B\beta_B,\label{aAbA}\\ 
d(*(*\theta^A\we\theta^B))&=\xi^Bd\xi^A+\xi^Ad\xi^B\label{dtt-dxx},
\end{align}
where in \eqref{ett-al} $\alpha$ is a one-form, and in \eqref{abg} $\alpha$ and $\beta$ are $k$-forms and $\gamma$ a one-form, in \eqref{a*bs_1} $\alpha$ and $\beta$ are $k$-forms and finally in \eqref{aAbA} $\alpha^A$ is a $k$-form, while $\beta_A$ is a $(3-k)$-form. Moreover, we will apply the following two identities: 
\begin{align}
\alpha\we*(*\gamma\we\beta)-*\alpha*(\beta\we\gamma)-\beta\we*(*\gamma\we\alpha)+*\beta*(\alpha\we\gamma)&=0,\label{agb}\\
*[\alpha\we*(\kappa\we\beta)]-\alpha*(\beta\we*\kappa)-*[\beta\we*(\kappa\we\alpha)]+\beta*(\alpha\we*\kappa)&=0,\label{akb}
\end{align}
where $\alpha,\beta$ and $\kappa$ are one-forms, and $\gamma$ is a two-form.
    
Note that if in \eqref{abg} $\alpha$ and $\beta$ are three-forms then the formula can be simplified further. Indeed, in this case $*\alpha$ and $*\beta$ are zero-forms thus
\[
(\vth_A\lr\alpha)\we*\beta\we\gamma=\alpha\we*\beta\vth_A\lr\gamma
\] 
and setting this equality to the r.h.s. of the first line of \eqref{abg} we obtain 
\begin{equation}
(\alpha\we\Star{A}\beta)\we\gamma=-\alpha\we*\beta\,\vth_A\lr\gamma.
\label{abg-3}
\end{equation}
Using this result we can also simplify \eqref{a*bs_1}---if $\alpha$ and $\beta$ are three-forms then setting to \eqref{abg-3} $\gamma=(\delta S_1(M))/(\delta p_A)$ we obtain  
\begin{equation}
(\alpha\we\Star{A}\beta)\we\frac{\delta S_1(M)}{\delta p_A}=\frac{M}{2}\alpha\we*\beta*(p_A\we\theta^A).
\label{a*bs_1-3}
\end{equation}

\begin{proof}[Proof of \eqref{vt-xi}.] Recall that the functions $\xi^B$ are given by the formula \eqref{xi-sol}. Using it we obtain   
\begin{multline*}
\vth_A\lr*\xi^B=-\frac{1}{2}\eps^{B}{}_{CDE}(\vth_A\lr\theta^C)\theta^D\we\theta^E=-\frac{1}{2}\eps^{B}{}_{ADE}\theta^D\we\theta^E-\frac{1}{2}\eps^B{}_{CDE}\xi_A\xi^C\theta^D\we\theta^E=\\=-\frac{1}{2}\eps^{B}{}_{CDA}\theta^C\we\theta^D+\xi_A*\theta^B,
\end{multline*}
where in the second step we used \eqref{tt-eta}, and in the last one \eqref{ett-xi}.
\end{proof}

\begin{proof}[Proof of \eqref{ett-al}.] By virtue of \eqref{tt-eta} and \eqref{ett-xi} the l.h.s. of \eqref{ett-al} can be transformed as follows:
\begin{multline*}
-\frac{1}{2}\eps^A{}_{BCD}\theta^B\we\theta^C\we\alpha=-\frac{1}{2}\eps^A{}_{BCE}\theta^B\we\theta^C\we(\vth_D\lr\theta^E-\xi_D\xi^E)\alpha=\\=-\frac{1}{2}\eps^A{}_{BCE}\theta^B\we\theta^C\we(\vth_D\lr\theta^E)\alpha-(*\theta^A)\we\xi_D\alpha.
\end{multline*}
Shifting the contraction $\vth_D\lr$ in the first of the two resulting terms and applying once again \eqref{tt-eta} and \eqref{ett-xi} we obtain
\begin{multline*}
-\frac{1}{2}\eps^A{}_{BCD}\theta^B\we\theta^C\we\alpha=\eps^A{}_{DCE}\theta^C\we\theta^E\we\alpha-\frac{1}{2}\eps^A{}_{BCE}\theta^B\we\theta^C\we\theta^E(\vth_D\lr\alpha)-\\-3(*\theta^A)\we\xi_D\alpha.
\end{multline*} 
Now to justify \eqref{ett-al} it is enough to note that $(i)$ the first term on the r.h.s of the equation above is proportional to the term on the l.h.s. and $(ii)$ the second term on the r.h.s. by virtue of \eqref{xi-sol} is equal to $3(*\xi^A)\vth_D\lr\alpha$.
\end{proof}

\begin{proof}[Proof of \eqref{abg}.] Let us now consider the l.h.s. of \eqref{abg}:
\begin{multline*}
(\alpha\Star{A}\beta)\we\gamma=\vth^B\lr\Big(\eta_{AB}\alpha\we*\beta-[(\vth_A\lr\alpha)\we*(\vth_B\lr\beta)+(\alpha\leftrightarrow\beta)]\Big)\we\gamma=\\=\alpha\we*\beta (\vth_A\lr\gamma)-[(\vth_A\lr\alpha)\we*(\vth_B\lr\beta)+(\alpha\leftrightarrow\beta)]\vth^B\lr\gamma.
\end{multline*}
By virtue of \eqref{a-b} and \eqref{tta=ka}
\[
*(\vth_B\lr\beta)\vth^B\lr\gamma=*\beta\we\theta_B\we\vth^B\lr\gamma=*\beta\we\gamma.
\]
Setting this result to the r.h.s. of the previous equation we obtain the r.h.s. of the first line of \eqref{abg}. To obtain the result at the second line it is enough to shift the contraction $\vth_A\lr$ in the term $-(\vth_A\lr\alpha)\we*\beta\we\gamma$.  
\end{proof}

\begin{proof}[Proof of \eqref{a*bs_1}.] By virtue of the first line of  Equation \eqref{abg} just proven 
\begin{multline}
(\alpha\we\Star{A}\beta)\we\frac{\delta S_1(M)}{\delta p_A}=\alpha\we*\beta\,\vth_A\lr\frac{\delta S_1(M)}{\delta p_A}-[(\vth_A\lr\alpha)\we*\beta+(\alpha\leftrightarrow\beta)]\we\frac{\delta S_1(M)}{\delta p_A}=\\=-\frac{M}{2}\alpha\we*\beta*(p_A\we\theta^A) -M[(\vth_A\lr\alpha)\we*\beta+(\alpha\leftrightarrow\beta)]\we\Big(\theta^B*(p_B\we\theta^A)-\frac{1}{2}\theta^A*(p_B\we\theta^B)\Big).
\label{a*bs_1-calc}
\end{multline}
Because $*(p_B\we\theta^A)$ is a zero-form (i.e. a function) 
\[
(\vth_A\lr\alpha)*(p_B\we\theta^A)=*(*\alpha\we\theta_A)*(p_B\we\theta^A)=*(*\alpha\we\theta_A\we*(p_B\we\theta^A))=*(*\alpha\we*p_B),
\]
where the first equality holds true by virtue of \eqref{a-b} and the last one---due to \eqref{t*at}. On the other hand due to \eqref{tta=ka} 
\[
(\vth_A\lr\alpha)\we*\beta\we\theta^A=\theta^A\we(\vth_A\lr\alpha)\we*\beta=k\alpha\we*\beta.
\]   
Setting the two results above to \eqref{a*bs_1-calc} we get 
\begin{multline}
(\alpha\we\Star{A}\beta)\we\frac{\delta S_1(M)}{\delta p_A}=M(k-\frac{1}{2})\alpha\we*\beta*(p_A\we\theta^A)-\\-M[*(*\alpha\we*p_A)\we*\beta\we\theta^A+(\alpha\leftrightarrow\beta)].
\label{a*bs_1-calc-1}
\end{multline}
The last step of the proof aims at simplifying the last term in the equation above:
\begin{multline*}
*(*\alpha\we*p_A)\we*\beta\we\theta^A=*\alpha\we*p_A\we*(*\beta\we\theta^A)=*\alpha\we*p_A\we\vth^A\lr\beta=\\=(-1)^{3-k}\vth^A\lr(*\alpha\we*p_A)\we\beta=(-1)^{3-k}*(\alpha\we\theta^A)\we*p_A\we\beta+*\alpha\we \beta*(p_A\we\theta^A)
\end{multline*}
(here we used \eqref{a-b} and \eqref{a-*g}). Taking into account that $*\alpha\we\beta=\alpha\we*\beta$ we set the result above to \eqref{a*bs_1-calc-1} obtaining thereby \eqref{a*bs_1}.  
\end{proof}

\begin{proof}[Proof of \eqref{aAbA}.] By virtue of  \eqref{tt-eta} 
\[
\alpha^A\we\beta_A=(\vth^B\lr\theta_A-\xi^B\xi_A)\alpha^A\we\beta_B=\vth^B\lr\theta_A \alpha^A\we\beta_B-\xi_A\alpha^A\we\xi^B\beta_B.
\]
Now to get \eqref{aAbA} it is enough to shift the contraction $\vth^B\lr$ in the first term at the r.h.s. of the equation above.   
\end{proof}

\begin{proof}[Proof of \eqref{dtt-dxx}] First transform the l.h.s. of \eqref{tt-eta} by means of \eqref{a-b} then act on the both sides of the resulting formula by $d$. 
\end{proof}

\begin{proof}[Proof of \eqref{agb}.] Note that $\beta\we\gamma$ is a three form. Therefore
\begin{equation*}
*\alpha*(\beta\we\gamma)=*[*(\beta\we\gamma)\we\alpha]=\vec{\alpha}\lr(\beta\we\gamma)=\vec{\alpha}\lr\beta\we\gamma-\beta\we\vec{\alpha}\lr\gamma,
\end{equation*}
where in the second step we used \eqref{a-b}. Transforming similarly the term $*\beta*(\alpha\we\gamma)$ we obtain 
\begin{multline*}
\alpha\we*(*\gamma\we\beta)-*\alpha*(\beta\we\gamma)-\beta\we*(*\gamma\we\alpha)+*\beta*(\alpha\we\gamma)=\\=\alpha\we\vec{\beta}\lr\gamma-\vec{\alpha}\lr\beta\we\gamma+\beta\we\vec{\alpha}\lr\gamma-\beta\we\vec{\alpha}\lr\gamma+\vec{\beta}\lr\alpha\we\gamma-\alpha\we\vec{\beta}\lr\gamma=0.
\end{multline*}
\end{proof}

\begin{proof}[Proof of \eqref{akb}.] Act by the Hodge operator $*$ on both sides of \eqref{agb} and set $\kappa=*\gamma$.
\end{proof}

Now we are ready to begin the calculations of $\{R(B),S(M)\}$. Let us recall that variations needed to calculate the bracket are given by \eqref{R-var} and \eqref{S1/t}--\eqref{S3/p}.

%***************************************************
\subsubsection{Terms quadratic in $p_A$ }
%***************************************************

Terms quadratic in the momenta come form the Poisson bracket
\begin{multline}
\{R_1(b),S_1(M)\}=\int_\Sigma[-b\we*p_A+(b\we\theta^B)\we\Star{A}p_B]\we\frac{\delta S_1(M)}{\delta p_A}-\\-M\Big(p_B*(p_A\we\theta^B)-\frac{1}{2}p_A*(p_B\we\theta^B)\Big)\we*(b\we\theta^A),
\label{r1-s1}
\end{multline}
where we used \eqref{abbt} to simplify the r.h.s. It is not difficult to see that
\[
-b\we*p_A\we\frac{\delta S_1(M)}{\delta p_A}-M\Big(p_B*(p_A\we\theta^B)-\frac{1}{2}p_A*(p_B\we\theta^B)\Big)\we*(b\we\theta^A)=0.
\] 
Let us now transform the remaining term in \eqref{r1-s1}---by virtue of \eqref{a*bs_1} 
\begin{multline*}
[(b\we\theta^B)\we\Star{A}p_B]\we\frac{\delta S_1(M)}{\delta p_A}=-\frac{M}{2}b\we\theta^B\we*p_B*(p_A\we\theta^A)+\\+M[*(b\we\theta^B\we\theta^A)\we*p_A\we p_B+*(p_B\we\theta^A)\we*p_A\we b\we\theta^B]=\\=-\frac{M}{2}b\we\theta^B\we*p_B*(p_A\we\theta^A)+M*(p_B\we\theta^A)*p_A\we b\we\theta^B.
\end{multline*}
To justify the last step let us note that $(b\we\theta^B\we\theta^A)$ is antisymmetric in $A$ and $B$ while $*p_A\we p_B$ is symmetric. Consequently,
\begin{equation}
\{R_1(b),S_1(M)\}=\int_\Sigma-\frac{M}{2}b\we\theta^B\we*p_B*(p_A\we\theta^A)+M*(p_B\we\theta^A) b\we\theta^B\we*p_A.
\label{RS-2}
\end{equation}

%***************************************************
\subsubsection{Terms linear in $p_A$}
%***************************************************

Here we will calculate the brackets $\{R_1(b),S_2(M)\}$ and $\{R_2(b),S_1(M)\}$, which give terms linear in the momenta. 

Considering $\{R_1(b),S_2(M)\}$ we immediately see that by virtue of \eqref{abbt} and \eqref{eps-bt}  
\[
\frac{\delta S_2(M)}{\delta \theta^A }\we\frac{\delta R_1(b)}{\delta p_A}=0,
\]
thus
\begin{multline}
\{R_1(b),S_2(M)\}=\int_\Sigma [-b\we *p_A+(b\we\theta^B)\we\Star{A}p_B]\we d(M\xi^A)=\int_\Sigma-dM\we b\we*(\xi^Ap_A)-\\-Mb\we*p_A\we d\xi^A+Mb\we\theta^B\we d\xi^A\vth_A\lr*p_B-M\vth_A\lr p_B\we*(b\we\theta^B)\we d\xi^A,
\label{R1S2}
\end{multline}
where we have used the second line of \eqref{abg} and \eqref{xi-d*=0}.

The other bracket,
\begin{equation}
\{R_2(b),S_1(M)\}=\int_\Sigma[d(b\xi_A)+(b\we d\theta^B)\we\Star{A}(*\xi_B)-\frac{1}{2}\eps_{DBCA}\theta^B\we\theta^C*(b\we d\theta^D)]\we \frac{\delta S_1(M)}{\delta p_A}.
\label{R2S1}
\end{equation}
The last two terms in the square bracket above give together zero once multiplied by $(\delta S_1(M))/(\delta p_A)$. Indeed, due to \eqref{a*bs_1-3} 
\[
[(b\we d\theta^B)\we\Star{A}(*\xi_B)]\we \frac{\delta S_1(M)}{\delta p_A}=\frac{M}{2}(b\we\xi_Bd\theta^B)*(p_A\we\theta^A).
\]
On the other hand, 
\begin{multline*}
-\frac{1}{2}\eps_{DBCA}\theta^B\we\theta^C\we \frac{\delta S_1(M)}{\delta p_A}*(b\we d\theta^D)=*\xi_D*(b\we d\theta^D)\vth_A\lr\frac{\delta S_1(M)}{\delta p_A}=\\=-\frac{M}{2}(b\we\xi_Bd\theta^B)*(p_A\we\theta^A)
\end{multline*}
---the first equality holds by virtue of \eqref{ett-al} and due to a fact that $\xi_A(\delta S_1(M))/(\delta p_A)=0$ ($(\delta S_1(M))/(\delta p_A)$ is of the form $\theta^A_i\gamma^i{}_jdx^j$ form some tensor field $\gamma^i{}_j$), in the last step we used \eqref{tt-eta}. Using the fact just mentioned and \eqref{a-*g} we transform the only remaining term in \eqref{R2S1} as follows:
\begin{multline}
d(b\xi_A)\we\frac{\delta S_1(M)}{\delta p_A}=-Mb\we d\xi_A\we \theta^B*(p_B\we\theta^A)+\frac{M}{2}b\we d\xi_A\we\theta^A*(p_B\we\theta^B)=\\=Mb\we \theta^B\we d\xi^A \vth_A\lr*p_B-\frac{M}{2}b\we\theta^A\we d\xi_A*(p_B\we\theta^B).
\label{R2S1-rem}
\end{multline}

Gathering all the terms linear in $p_A$, that is, \eqref{R1S2} and \eqref{R2S1-rem} we obtain
\begin{multline}
\{R_1(b),S_2(M)\}-\{R_2(b),S_1(M)\}=\int_\Sigma-dM\we b\we*(\xi^Ap_A)-Mb\we*p_A\we d\xi^A-\\-M\vth_A\lr p_B\we*(b\we\theta^B)\we d\xi^A+\frac{M}{2}b\we \theta^A\we d\xi_A*(p_B\we\theta^B).
\label{RS-1}
\end{multline} 

%***************************************************
\subsubsection{Terms independent of $p_A$ }
%***************************************************

It turns out that the remaining three brackets, $\{R_2(b),S_3(M)\}$, $\{R_2(b),S_2(M)\}$ and $\{R_1(b),S_3(M)\}$ do not depend on the momenta. The first bracket is zero since both $R_2(b)$ and $S_3(M)$ do not contain $p_A$. The second bracket contains a term $d(b\xi_A)\we d(M\xi^A)$ being an exact three-form---the integral of this term over $\Sigma$ is zero hence  
\begin{multline*}
\{R_2(b),S_2(M)\}=\int_\Sigma[(b\we d\theta^B)\we\Star{A}(*\xi_B)-\frac{1}{2}\eps_{DBCA}\theta^B\we\theta^C*(b\we d\theta^D)]\we d(M\xi^A).
\end{multline*}
Applying \eqref{abg-3}, \eqref{ett-xi} and \eqref{ett-al} it is not difficult to show that
\[
\{R_2(b),S_2(M)\}=\int_\Sigma dM\we*\theta_D*(b\we d\theta^D).
\]

Consider now the last bracket $\{R_1(b),S_3(M)\}$. Due to \eqref{abbt}
\begin{multline*}
\{R_1(b),S_3(M)\}=-\int_\Sigma \Big(d[M\theta^B*(d\theta_B\we\theta_A)-\frac{M}{2}\theta_A*(d\theta_B\we\theta^B)]+Md\theta_B*(d\theta_A\we\theta^B)-\\-\frac{M}{2}d\theta_A*(d\theta_B\we\theta^B)\Big)\we*(b\we\theta^A). 
\end{multline*}     
Our goal now is to express the r.h.s. of the equation above as a sum of a term containing $Mb$ and a one containing $dM$. To this end we first act by the operator $d$ on the factors constituting terms in the square brackets. Next, in those cases when it is possible, we use \eqref{t*at} to simplify $\theta_A\we*(b\we\theta^A)$ to $2*b$ and $\theta_A\we*(d\theta_B\we\theta^A)$ to $*d\theta_B$, finally in all the terms containing $M$ and $*b$ we shift the Hodge operator $*$ to get $b$. Thus we obtain
\begin{multline*}
\{R_1(b),S_3(M)\}=-\int_\Sigma Mb\we[-*d\theta^A\we*d\theta_A+\theta^A\we*d\theta^B*(d\theta_A\we\theta_B)-\\-\theta^A\we*d\theta_A*(d\theta_B\we\theta^B)-\theta^A\we*(\theta^B\we d*(d\theta_B\we\theta_A))-*d*(d\theta_A\we\theta^A)]+\\+dM\we[\theta^A\we*(b\we*d\theta_A)-*b*(d\theta_A\we\theta^A)].
\end{multline*}
Note that since $*d\theta^A$ is a one-form the first term at the r.h.s. above vanishes. Thus the terms independent of $p_A$ read
\begin{multline*}
\{R_1(b),S_3(M)\}-\{R_2(b),S_2(M)\}=\int_\Sigma \text{the term containing $Mb$}\,-\\-\int_\Sigma dM\we[\theta^B\we*(b\we*d\theta_B)-*b*(d\theta_B\we\theta^B)+*\theta_D*(b\we d\theta^D)]. 
\end{multline*}
The term containing $dM$ can be expressed as  
\[
-\int_\Sigma dM\we[-*b*(\theta^B\we d\theta_B)-\theta^B\we*(*d\theta_B\we b)+*\theta^B*(b\we d\theta_B)].
\]
Now by setting in \eqref{agb} $\alpha=b$, $\beta=\theta^B$ and $\gamma=d\theta_B$ the term under consideration can be simplified to
\[
-\int_\Sigma dM\we[-b\we*(*d\theta_B\we\theta^B)]=-\int_\Sigma*(dM\we b)\we\theta^B\we *d\theta_B.
\]   
This means that the terms independent of $p_A$ read 
\begin{multline}
\{R_1(b),S_3(M)\}-\{R_2(b),S_2(M)\}=-\int_\Sigma Mb\we[\theta^A\we*d\theta^B*(d\theta_A\we\theta_B)-\\-\theta^A\we*d\theta_A*(d\theta_B\we\theta^B)-\theta^A\we*(\theta^B\we d*(d\theta_B\we\theta_A))-*d*(d\theta_A\we\theta^A)]-\\-\int_\Sigma*(dM\we b)\we\theta^A\we *d\theta_A.
\label{RS-0}
\end{multline}

%***************************************************
\subsubsection{Isolating constraints}
%***************************************************

Our goal now is to express the bracket 
\begin{multline*}
\{R(b),S(M)\}= \text{the terms \eqref{RS-2} quadratic in $p_A$}+\\+\text{the terms \eqref{RS-1} linear in $p_A$}+\text{the terms \eqref{RS-0} independent of $p_A$ }
\end{multline*}
as a sum of the constraints \eqref{B-sm}--\eqref{V-sm} smeared with some fields. 

\paragraph{Terms quadratic in $p_A$} We are going to transform the last term of \eqref{RS-2} to a form containing the factor $\theta^A\we*p_A$ being a part of the constraint $R(b)$: applying \eqref{aAbA} to the term with $\alpha^A=*p^A$ we obtain  
\begin{multline}
\Big(M*(p_B\we\theta^A)\we b\we\theta^B\Big)\we*p_A=*p^A\we\Big(M*(p_C\we\theta_A)\we b\we\theta^C\Big)=\\=\theta_A\we\vth^B\lr*p^A\we\Big(M*(p_C\we\theta_B)\we b\we\theta^C\Big)-\theta_A\we*p^A\we\vth^B\lr\Big(M*(p_C\we\theta_B)\we b\we\theta^C\Big).
\label{M*pB}
\end{multline}
The first term in the last line is zero---indeed, using \eqref{a-*g} we get
\[
\theta_A\we\vth^B\lr*p^A\we\Big(M*(p_C\we\theta_B)\we b\we\theta^C\Big)=-M*(p_A\we\theta^B)*(p_C\we\theta_B)\theta^A\we\theta^C\we b.
\] 
Note now that $*(p_A\we\theta^B)*(p_C\we\theta_B)$ is symmetric in $A$ and $C$, while $\theta^A\we\theta^C$ antisymmetric. Transforming the remaining term in \eqref{M*pB} we obtain
\begin{multline*}
\Big(M*(p_B\we\theta^A)\we b\we\theta^B\Big)\we*p_A=\\=-M\theta_A\we*p^A\we\Big(*(p_C\we\theta_B\,\vth^B\lr b)\we\theta^C-*(p_C\we\theta^B)b\,\vth^B\lr\theta^C\Big)=\\=-M\theta^A\we*p_A\we\Big(*(p_C\we b)\theta^C-*(p_B\we\theta^B)b\Big).
\end{multline*}
where in the last step we used \eqref{tta=ka} and \eqref{tt-eta}. 

Setting this result to \eqref{RS-2} we arrive at
\begin{multline}
\text{the terms \eqref{RS-2} quadratic in $p_A$}=-\int_\Sigma M[\theta^B*(p_B\we b)-\frac{1}{2}b*(p_B\we\theta^B)]\we\theta^A\we*p_A=\\=-R\Big(M[\theta^B*(p_B\we b)-\frac{1}{2}b*(p_B\we\theta^B)]\Big)-\int_\Sigma M[\theta^B*(p_B\we b)-\frac{1}{2}b*(p_B\we\theta^B)]\we\theta^A\we d\xi_A.
\label{RS-2a}
\end{multline}

Consequently, the bracket $\{R(b),S(M)\}$ is of the following form
\begin{multline}
\{R(b),S(M)\}=-R\Big(M[\theta^B*(p_B\we b)-\frac{1}{2}b*(p_B\we\theta^B)]\Big)+\text{terms linear in $p_A$} +\\+\text{the terms \eqref{RS-0} independent of $p_A$},
\label{RS-1-0}
\end{multline}
where now the phrase ``terms linear in $p_A$'' means the terms \eqref{RS-1} and the last term in \eqref{RS-2a}. Now we are going to isolate constraints from the linear terms.

\paragraph{Terms linear in $p_A$} The terms read
\begin{multline}
\int_\Sigma-dM\we b\we*(\xi^Ap_A)+Mb\we d\xi^A\we*p_A-M\vth_A\lr p_B\we*(b\we\theta^B)\we d\xi^A+\\+Mb\we\theta^A\we d\xi_A*(p_B\we\theta^B)-M\theta^B\we\theta^A\we d\xi_A*(p_B\we b).
\label{RS-1a}
\end{multline}
The first term can be written as 
\begin{equation}
\int_\Sigma-dM\we b\we*(\xi^Ap_A)=\int_\Sigma-*(dM\we b)\we\xi^Ap_A=-B(*(dM\we b))+\int_\Sigma *(dM\we b)\we\theta^A\we *d\theta_A.
\label{RS-1a+dM}
\end{equation}

Transformation of the remaining terms in \eqref{RS-1a} (i.e. those which do not contain $dM$) to an appropriate form takes more effort. Applying \eqref{aAbA} to the first of them by setting $\alpha^A=b\we d\xi^A$ and  $\beta_A=*p_A$ we obtain
\begin{multline*}
Mb\we d\xi^A\we*p_A=M\theta_A\we\vth^B\lr(b\we d\xi^A)\we *p_B+M\theta_A\we b\we d\xi^A\vth^B\lr*p_B=\\=M\theta_A\we d\xi^A\we *p_B\,\vth^B\lr b- M\theta_A\we b\we *p_B\,\vth^B\lr d\xi^A-M b\we \theta_A\we d\xi^A *(p_B\we\theta^B)
\end{multline*}  
---here in the last step we used \eqref{a-*g}. The last term of \eqref{RS-1a} can be transformed as follows
\begin{multline*}
-M\theta^B\we\theta^A\we d\xi_A*(p_B\we b)=-M\theta^A\we d\xi_A\we**[*(p_B\we b)\we\theta^B]=\\=-M\theta^A\we d\xi_A\we*[\vth^B\lr(p_B\we b)]=
-M\theta^A\we d\xi_A\we*[(\vth^B\lr p_B) \we b]-M\theta^A\we d\xi_A\we*p_B\,\vth^B\lr b,
\end{multline*}
where in the second step we used \eqref{a-b}. The last two results allow us to express in a more simpler form the sum of the terms in \eqref{RS-1a} which do not contain $dM$:
\begin{equation}
- M\theta_A\we b\we *p_B\,\vth^B\lr d\xi^A-M\vth_A\lr p_B\we*(b\we\theta^B)\we d\xi^A-M\theta^A\we d\xi_A\we*[(\vth^B\lr p_B) \we b].
\label{RS-1a-dM}
\end{equation}

Our goal now is to rewrite the sum above in a form of a single term containing the factor $\theta^A\we*p_A$. Let us begin with the first term in \eqref{RS-1a-dM}: 
\[
- M\theta_A\we b\we *p_B\,\vth^B\lr d\xi^A=M\Big(*(b\we\theta_C)\,\vth^A\lr d\xi^C\Big)\we p_A.
\]
Setting in \eqref{aAbA} $\alpha^A=*(b\we\theta_C)\,\vth^A\lr d\xi^C$ and $\beta_A=p_A$ we obtain
\begin{multline}
- M\theta_A\we b\we *p_B\,\vth^B\lr d\xi^A=\\=M\theta_A\we\vth_B\lr\Big(*(b\we\theta_C)\,\vth^A\lr d\xi^C\Big)\we p_B-M\theta_A\Big(*(b\we\theta_C)\,\vth^A\lr d\xi^C\Big)\we\vth^B\lr p_B=\\=M\theta_A(\vth^A\lr d\xi^C)\we*(b\we\theta_C\we\theta^B)\we p_B-M\theta_A(\vth^A\lr d\xi^C)\we*(b\we\theta_C)\we *(*p_B\we\theta^B)=\\=
Md\xi^C \we*(b\we\theta_C\we\theta^B)\we p_B+M*[d\xi^C\we*(b\we\theta_C)]\we\theta^B\we*p_B,
\label{RS-1a-dM-1}
\end{multline}
 where we applied \eqref{a-*g} and \eqref{a-b} in the third step and \eqref{tta=ka} in the last step. The second term in \eqref{RS-1a-dM} 
\begin{multline}
-M\vth_A\lr p_B\we*(b\we\theta^B)\we d\xi^A=Mp_B [\vth_A\lr*(b\we\theta^B)]\we d\xi^A-Mp_B\we*(b\we\theta^B)\,\vth_A\lr d\xi^A=\\=Mp_B*(b\we\theta^B\we\theta_A)\we d\xi^A-M\vth_C\lr d\xi^C*(b\we\theta^A)\we p_A=\\=-Md\xi^C*(b\we\theta_C\we\theta^B)\we p_B-M\vth_C\lr d\xi^C b\we \theta^B\we* p_B
\label{RS-1a-dM-2}
\end{multline}
(in the second step we applied \eqref{a-*g}). Finally, the last term in \eqref{RS-1a-dM} by virtue of \eqref{a-b} can be written as
\begin{multline*}
-M\theta^A\we d\xi_A\we*[(\vth^B\lr p_B) \we b]=-M*(*p_B\we\theta^B)\we b\we*(\theta^A\we d\xi_A)=\\M*[b\we*(\theta^A\we d\xi_A)]\we\theta^B\we*p_B.
\end{multline*}
Gathering \eqref{RS-1a-dM-1}, \eqref{RS-1a-dM-2} and the equation above we obtain the desired expression for these terms in \eqref{RS-1a} which do not contain $dM$:
\begin{multline*}
\int_\Sigma M\Big(*[d\xi^C\we*(b\we\theta_C)]-\vth_C\lr d\xi^C b+*[b\we*(\theta^A\we d\xi_A)]\Big)\we\theta^B\we*p_B=\\=\int_\Sigma M\Big(*[b\we*(\theta^A\we d\xi_A)]-b*(d\xi_A\we*\theta^A)-*[d\xi_A\we*(\theta^A\we b)]\Big)\we\theta^B\we*p_B, 
\end{multline*}
where again we used \eqref{a-b}. We can now simplify the term in the big parenthesis --- it is enough to use \eqref{akb} setting $\alpha=b$, $\beta=d\xi_A$ and $\kappa=\theta^A$ to get 
\begin{multline}
\int_\Sigma M\Big(-d\xi_A*(b\we*\theta^A)\Big)\we\theta^B\we*p_B=-R\Big(Md\xi_A*(b\we*\theta^A)\Big)-\\-\int_\Sigma M*(b\we*\theta^A)d\xi_A\we\xi^B d\theta_B,
\label{RS-1a-dM-fn}
\end{multline}
which is the final form of the terms in \eqref{RS-1a} which do not contain $dM$.

The equations \eqref{RS-1a+dM} and \eqref{RS-1a-dM-fn} allow us to express the terms linear in $p_A$ appearing in \eqref{RS-1-0} as a sum of constraints and terms independent of the momenta. Consequently, \eqref{RS-1-0} can be written in the following form:
\begin{multline}
\{R(b),S(M)\}=-R\Big(M[\theta^B*(p_B\we b)-\frac{1}{2}b*(p_B\we\theta^B)+d\xi_A*(b\we*\theta^A)]\Big)-B\Big(*(dM\we b)\Big)+\\+\text{terms independent of $p_A$},
\label{RS-c-0}
\end{multline}
where the phrase ``terms independent of $p_A$'' means here the terms given by \eqref{RS-0}, the last term in \eqref{RS-1a+dM} and the last one in \eqref{RS-1a-dM-fn}.

\paragraph{Terms independent of $p_A$} Our goal now is to show that the terms independent of $p_A$ sum up to zero. Gathering all the terms under consideration which appear in \eqref{RS-c-0} we see that the sum of the last term of \eqref{RS-0} and the last term  of \eqref{RS-1a+dM} is zero. Note now that the last term in \eqref{RS-1a-dM-fn} contains $\xi^A$ which does not appear in the others term. To get rid of $\xi^A$ let us use \eqref{dtt-dxx}:   
\begin{multline*}
-\int_\Sigma M*(b\we*\theta^A)d\xi_A\we\xi^B d\theta_B=-\int_\Sigma M*(b\we*\theta^A)d*(*\theta_A\we\theta^B)\we d\theta_B=\\=-\int_\Sigma Mb\we*\theta^A*[d*(*\theta_A\we\theta^B)\we d\theta_B].
\end{multline*}
Consequently, the terms in \eqref{RS-c-0} independent of $p_A$ read
\begin{multline}
-\int_\Sigma Mb\we\Big(\theta^A\we*d\theta^B*(d\theta_A\we\theta_B)-\theta^A\we*d\theta_A*(d\theta_B\we\theta^B)-\theta^A\we*(\theta^B\we d*(d\theta_B\we\theta_A))-\\-*d*(d\theta_A\we\theta^A)+*\theta^A*(d*(*\theta_A\we\theta^B)\we d\theta_B)\Big).
\label{mb-ttt}
\end{multline}

Now we are going to show that the expression above is zero for every $M,b$ and $\theta^A$. This will be achieved by proving that the terms in the big parenthesis sum up to zero for every $\theta^A$. The proof will be carried out with application of tensor calculus (see formulae in Section \ref{t-calc}).
  
The first term in \eqref{mb-ttt}
\[
\theta^A\we*d\theta^B*(d\theta_A\we\theta_B)=(\nabla_i\theta_{Aj})(\nabla^a\theta^{Bb})\theta_{Bk}\theta^A_d\eps_{abc}\eps^{ijk}\,dx^d\we dx^c.
\]
By virtue of the third equation in \eqref{eps-delta} we can express the r.h.s. above as a sum of six terms. Four of them vanish: two of them contain the vanishing factor $(\nabla_a\theta^{Bb})\theta_{Bb}$ (see \eqref{nab-tt}), the remaining two vanish because they are of the form 
\begin{equation}
\gamma_{dc}\,dx^d\we dx^c, \ \ \gamma_{dc}=\gamma_{cd}. 
\label{al-sym}
\end{equation} 
Thus
\begin{multline}
\theta^A\we*d\theta^B*(d\theta_A\we\theta_B)=(\nabla_b\theta_{Ac})(\nabla^a\theta^{Bb})\theta_{Ba}\theta^A_d\,dx^d\we dx^c-\\-(\nabla_c\theta_{Ab})(\nabla^a\theta^{Bb})\theta_{Ba}\theta^A_d\,dx^d\we dx^c.
\label{ttt-1}
\end{multline}

In the case of the second term in \eqref{mb-ttt} we proceed similarly---applying \eqref{eps-delta} we obtain six terms and again four of them vanish: two of them are of the form \eqref{al-sym}, the other two can be transformed to this form by means of \eqref{0-nab-q} hence 
\begin{multline}
-\theta^A\we*d\theta_A*(d\theta_B\we\theta^B)=-(\nabla^a\theta^{b}_A)(\nabla_c\theta^{B}_a)\theta_{Bb}\theta^A_d\,dx^d\we dx^c+\\+(\nabla^a\theta^b_A)(\nabla_c\theta^{B}_b)\theta_{Ba}\theta^A_d\,dx^d\we dx^c.
\label{ttt-2}
\end{multline}

Let us consider now the third term in \eqref{mb-ttt}:
\begin{multline*}
-\theta^A\we*(\theta^B\we d*(d\theta_B\we\theta_A))=-\nabla^b[(\nabla_i\theta_{Bj})\theta_{Ak}]\theta^{Ba}\theta^A_d\eps_{abc}\eps^{ijk}\,dx^d\we dx^c=\\=-(\nabla^b\nabla_i\theta_{Bj})\theta^{Ba}q_{kd}\eps_{abc}\eps^{ijk}\,dx^d\we dx^c-(\nabla_i\theta_{Bj})(\nabla^b\theta_{Ak})\theta^{Ba}\theta^A_d\eps_{abc}\eps^{ijk}\,dx^d\we dx^c,
\end{multline*}
where in the second step we used \eqref{q}. Applying \eqref{eps-delta} we obtain twelve terms: six of them contain a second covariant derivative of components of $\theta^B$, while the remaining ones are quadratic in covariant derivatives of the components. Three terms of those containing second covariant derivatives vanish: two terms turn out to be of the form \eqref{al-sym}, the third one can be transformed to this form by means of \eqref{nab-tt}:
\begin{multline*}
-(\nabla_d\nabla_c\theta_{Ba})\theta^{Ba}\,dx^d\we dx^c=-\nabla_d[(\nabla_c\theta_{Ba})\theta^{Ba}]+(\nabla_c\theta_{Ba})(\nabla_d\theta^{Ba})\,dx^d\we dx^c=\\=(\nabla_c\theta_{Ba})(\nabla_d\theta^{Ba})\,dx^d\we dx^c.
\end{multline*}
Four terms of those quadratic in covariant derivatives are zero: two of them contain the factor \eqref{nab-tt}, the other two can be transformed to the form \eqref{al-sym} by means of \eqref{0-nab-q}. Finally
\begin{multline}
-\theta^A\we*(\theta^B\we d*(d\theta_B\we\theta_A))=-(\nabla^b\nabla_b\theta_{Bc})\theta^{B}_d\,dx^d\we dx^c+(\nabla_d\nabla_a\theta_{Bc})\theta^{Ba}\,dx^d\we dx^c+\\+(\nabla^b\nabla_c\theta_{Bb})\theta^{B}_d\,dx^d\we dx^c-(\nabla_a\theta_{Bb})(\nabla^b\theta_{Ac})\theta^{Ba}\theta^A_d\,dx^d\we dx^c+\\+(\nabla_c\theta_{Bb})(\nabla^b\theta_{Aa})\theta^{Ba}\theta^A_d\,dx^d\we dx^c.
\label{ttt-3}
\end{multline}

Since
\[
d\theta_A\we\theta^A=3!(\nabla_{[d} \theta_{Ac})\theta^A_{b]}\,dx^d\ot dx^c\ot dx^b
\]
we can use \eqref{delta-b} to express the fourth term in \eqref{mb-ttt} as follows
\begin{multline*}
-*d*(d\theta_A\we\theta^A)=-\frac{3!}{2}\nabla^b[(\nabla_{[d} \theta_{Ac})\theta^A_{b]}]\,dx^d\we dx^c=-\nabla^b[(\nabla_{d} \theta_{Ac})\theta^A_{b}]\,dx^d\we dx^c-\\-\nabla^b[(\nabla_{c} \theta_{Ab})\theta^A_{d}]\,dx^d\we dx^c-\nabla^b[(\nabla_{b} \theta_{Ad})\theta^A_{c}]\,dx^d\we dx^c.
\end{multline*}
Acting by $\nabla^b$ on the factors in the square brackets we obtain
\begin{multline}
-*d*(d\theta_A\we\theta^A)=-(\nabla^b\nabla_{d} \theta_{Ac})\theta^A_{b}\,dx^d\we dx^c-(\nabla^b\nabla_{c} \theta_{Ab})\theta^A_{d}\,dx^d\we dx^c-\\-(\nabla^b\nabla_{b} \theta_{Ad})\theta^A_{c}\,dx^d\we dx^c-(\nabla_{d} \theta_{Ac})(\nabla^b\theta^A_{b})\,dx^d\we dx^c-(\nabla_{c} \theta_{Ab})(\nabla^b\theta^A_{d})\,dx^d\we dx^c,
\label{ttt-4}
\end{multline} 
where we omitted the term $-(\nabla_{b} \theta_{Ad})(\nabla^b\theta^A_{c})\,dx^d\we dx^c$ which being of the form \eqref{al-sym} is zero.

The last term in \eqref{mb-ttt} by virtue of \eqref{a-b} and \eqref{eps-delta} can be expressed as follows
\begin{multline*}
*\theta^A*(d*(*\theta_A\we\theta^B)\we d\theta_B)=\frac{1}{2}\theta^{Ab}\nabla_i(\theta_{Aa}\theta^{Ba})(\nabla_j\theta_{Bk})\eps^{ijk}\eps_{bdc}\,dx^d\we dx^c=\\=\theta^{Ab}\nabla_b(\theta_{Aa}\theta^{Ba})(\nabla_d\theta_{Bc})\,dx^d\we dx^c+\theta^{Ab}\nabla_d(\theta_{Aa}\theta^{Ba})(\nabla_c\theta_{Bb})\,dx^d\we dx^c+\\+\theta^{Ab}\nabla_c(\theta_{Aa}\theta^{Ba})(\nabla_b\theta_{Bd})\,dx^d\we dx^c.
\end{multline*}   
The last term in the second line above vanishes---indeed, due to \eqref{q} the term is equal to 
\[
\theta^{Ab}(\nabla_d\theta_{Aa})\theta^{Ba}(\nabla_c\theta_{Bb})\,dx^d\we dx^c +(\nabla_d\theta^{Bb})(\nabla_c\theta_{Bb})\,dx^d\we dx^c
\]
and each term in this sum is of the form \eqref{al-sym}. Using \eqref{q} again we obtain
\begin{multline}
*\theta^A*(d*(*\theta_A\we\theta^B)\we d\theta_B)=(\nabla_b\theta^{Bb})(\nabla_d\theta_{Bc})\,dx^d\we dx^c+(\nabla_c\theta^{Bb})(\nabla_b\theta_{Bd})\,dx^d\we dx^c+\\+(\nabla_b\theta_{Aa})(\nabla_d\theta_{Bc})\theta^{Ab}\theta^{Ba}\,dx^d\we dx^c+(\nabla_c\theta_{Aa})(\nabla_b\theta_{Bd})\theta^{Ab}\theta^{Ba}\,dx^d\we dx^c.
\label{ttt-5}
\end{multline}

Now we are ready to gather all the results \eqref{ttt-1}--\eqref{ttt-5} to show that \eqref{mb-ttt} is zero. To make the task easier let us note that we obtained three kinds of terms: $(i)$ ones containing second covariant derivatives of $\theta^A_a$, $(ii)$ ones quadratic in covariant derivatives of $\theta^A_a$ and $(iii)$ ones quadratic both in the covariant derivatives and in $\theta^A_a$. 

Expressions containing second covariant derivatives of $\theta^A_a$ appear in \eqref{ttt-3} and \eqref{ttt-4} and they read
\begin{multline*}-(\nabla^b\nabla_b\theta_{Bc})\theta^{B}_d\,dx^d\we dx^c+(\nabla_d\nabla_a\theta_{Bc})\theta^{Ba}\,dx^d\we dx^c+(\nabla^b\nabla_c\theta_{Bb})\theta^{B}_d\,dx^d\we dx^c-\\-(\nabla^b\nabla_{d} \theta_{Ac})\theta^A_{b}\,dx^d\we dx^c-(\nabla^b\nabla_{c} \theta_{Ab})\theta^A_{d}\,dx^d\we dx^c-(\nabla^b\nabla_{b} \theta_{Ad})\theta^A_{c}\,dx^d\we dx^c
\end{multline*}   
We see now that the first and the last term sum up to zero, similarly do the third and the fifth ones. The sum of the remaining second an fourth terms can be expressed as
\begin{multline*}
[(\nabla_d\nabla_a-\nabla_a\nabla_d)\theta^b_{B}]\theta^{Ba}q_{bc}\,dx^d\we dx^c=R^b{}_{eda}\theta^e_B\theta^{Ba}q_{bc}\,dx^d\we dx^c=\\=R^b{}_{eda}q^{ea}q_{bc}\,dx^d\we dx^c=R_{ceda}q^{ea}\,dx^d\we dx^c=R_{cd}\,dx^d\we dx^c=0,
\end{multline*}
where $(i)$ in the first step we used the Riemann tensor $R^b{}_{eda}$ of the Levi-Civita connection compatible with $q$ to express the commutator $(\nabla_d\nabla_a-\nabla_a\nabla_d)$ acting on $\theta^b_B$ and $(ii)$ in the second step we applied \eqref{q}. Note that the last equality holds by virtue of symmetricity of the Ricci tensor $R_{cd}$.    

Terms quadratic in covariant derivatives of $\theta^A_a$ appear in \eqref{ttt-4} and \eqref{ttt-5}:
\begin{multline*}
-(\nabla_{d} \theta_{Ac})(\nabla^b\theta^A_{b})\,dx^d\we dx^c-(\nabla_{c} \theta_{Ab})(\nabla^b\theta^A_{d})\,dx^d\we dx^c+(\nabla_b\theta^{Bb})(\nabla_d\theta_{Bc})\,dx^d\we dx^c+\\+(\nabla_c\theta^{Bb})(\nabla_b\theta_{Bd})\,dx^d\we dx^c
\end{multline*}  
It is easy to see that the first and the third terms sum up to zero, similarly do the second and fourth ones.

Let us finally consider the terms quadratic in covariant derivatives of $\theta^A_a$ and quadratic in $\theta^A_a$---there are eight of them and they can be grouped into pairs such that each pair sums up to zero. These pairs are: 
\begin{enumerate}
\item  the first term at the r.h.s. of \eqref{ttt-1} and the fourth one at the r.h.s. of \eqref{ttt-3},
\item the second term at the r.h.s. of \eqref{ttt-1} and the third one at the r.h.s. of \eqref{ttt-5} (apply \eqref{0-nab-q} to the latter term),
\item the first term at the r.h.s. of \eqref{ttt-2} and the fifth one at the r.h.s. of \eqref{ttt-3},
\item the second term at the r.h.s. of \eqref{ttt-2} and the fourth one at the r.h.s. of \eqref{ttt-5} (apply \eqref{0-nab-q} to the latter term).
\end{enumerate} 

In this way we demonstrated that \eqref{mb-ttt} is zero for every $M$, $b$ and $\theta^A$. Thus the formula \eqref{RS-c-0} turns into  the final expression of the Poisson bracket of $R(b)$ and $S(M)$:
\begin{equation}
\{R(b),S(M)\}=-R\Big(M[\theta^B*(p_B\we b)-\frac{1}{2}b*(p_B\we\theta^B)+d\xi_A*(b\we*\theta^A)]\Big)-B\Big(*(dM\we b)\Big).
\label{RS-fin}
\end{equation}

%***************************************************
\subsection{Poisson bracket of $B(a)$ and $S(M)$}
%***************************************************

The bracket 
\begin{equation}
\{B(a),S(M)\}=\sum_{i=1}^2\sum_{j=1}^3\{B_i(a),S_j(M)\}
\label{B-S}
\end{equation}
will be calculated in a similar way to $\{R(b),S(M)\}$. Recall that the constraints under consideration are defined by \eqref{B-sm} and \eqref{S-sm}, the functionals appearing at the r.h.s. of \eqref{B-S} are given by \eqref{B1B2} and \eqref{S_i} and that formulae \eqref{B-var} and \eqref{S1/t}--\eqref{S3/p} describe variations needed to calculate the bracket.

%***************************************************
\subsubsection{Terms quadratic in $p_A$ }
%***************************************************

The only term in \eqref{B-S} quadratic in $p_A$ is 
\begin{equation}
\{B_2(a),S_1(M)\}=\int_\Sigma \frac{\delta B_2(a)}{\delta\theta^A}\we\frac{\delta S_1(M)}{\delta p_A}-\frac{\delta S_1(M)}{\delta\theta^A}\frac{\delta B_2(a)}{\delta p_A}.
\label{B2-S1}
\end{equation}

The first term of the r.h.s. of this equation turns out to be zero. To show this let us express the term as follows
\begin{multline}
\frac{\delta B_2(a)}{\delta\theta^A}\we\frac{\delta S_1(M)}{\delta p_A}=-\frac{M}{2}\eps^D{}_{BCA}\theta^B\we\theta^C\we\theta^E*(a\we p_D)*(p_E\we\theta^A)+\\+\frac{M}{4}\eps^D{}_{BCA}\theta^B\we\theta^C\we\theta^A *(a\we p_D)*(p_E\we\theta^E)+[(*\xi^D)\we\Star{A}(a\we p_D)]\we\frac{\delta S_1(M)}{\delta p_A}.
\label{B2-S1=0}
\end{multline}
Our strategy now is to restore in each term above the function $\xi^A$ which originally appears in $B_2(a)$. Thus by virtue of \eqref{ett-al}  
\begin{multline}
-\frac{M}{2}\eps^D{}_{BCA}\theta^B\we\theta^C\we\theta^E*(a\we p_D)*(p_E\we\theta^A)=M(*\xi^D)(\vth_A\lr\theta^E)*(a\we p_D)*(p_E\we\theta^A)=\\=Ma\we\xi^Dp_D*(p_A\we\theta^A).
\label{BS-2-1}
\end{multline}
Due to \eqref{xi-sol} the second term at the r.h.s of \eqref{B2-S1=0} 
\begin{equation}
\frac{M}{4}\eps^D{}_{BCA}\theta^B\we\theta^C\we\theta^A\we*(a\we p_D)*(p_E\we\theta^E)=-\frac{3M}{2}a\we\xi^Dp_D*(p_A\we\theta^A).
\label{BS-2-2}
\end{equation}
By virtue \eqref{a*bs_1-3} 
\begin{equation}
[(*\xi^D)\we\Star{A}(a\we p_D)]\we\frac{\delta S_1(M)}{\delta p_A}=\frac{M}{2}a\we\xi^Dp_D*(p_A\we\theta^A).
\label{BS-2-3}
\end{equation}
Gathering the three results \eqref{BS-2-1}, \eqref{BS-2-2} and \eqref{BS-2-3} we see that, indeed, the first term on the r.h.s. of \eqref{B2-S1} is zero.

The second term at the r.h.s. of \eqref{B2-S1} requires \eqref{xi-d*=0} to be applied, then some simple transformations give us an expression for terms in \eqref{B-S} quadratic in the momenta:
\begin{equation}
\{B_2(a),S_1(M)\}=-\int_\Sigma M[\theta^B*(p_B\we a)-\frac{1}{2}a*(p_B\we\theta^B)]\we\xi^Ap_A.
\label{BS-2}
\end{equation}

%***************************************************
\subsubsection{Terms linear in $p_A$ }
%***************************************************

It turns out that $\{B_1(a),S_1(M)\}$ and $\{B_2(a),S_2(M)\}$ give terms linear in $p_A$. The first of the two brackets can be calculated as follows
\begin{multline*}
\{B_1(a),S_1(M)\}=\int_\Sigma[-a\we*d\theta_A+d*(a\we\theta_A)]\we M[\theta^B*(p_B\we\theta^A)-\frac{1}{2}\theta^A*(p_B\we\theta^B)]+\\+
[(a\we\theta^B)\we\Star{A}d\theta_B]\we\frac{\delta S_{1}(M)}{\delta p_A}
\end{multline*}
Using \eqref{a*bs_1} after some simple algebra we obtain
\begin{multline*}
\{B_1(a),S_1(M)\}=\int_\Sigma Ma\we\theta^B\we*d\theta_A*(p_B\we\theta^A)+Md*(a\we\theta_A)\we\theta^B*(p_B\we\theta^A)-\\-Ma\we\theta^B\we*d\theta_B*(p_A\we\theta^A)-\frac{M}{2}d*(a\we\theta_A)\we\theta^A*(p_B\we\theta^B)+\\+M*(a\we\theta^B\we\theta^A)*p_A\we d\theta_B-Ma\we*p_A\we\theta^B*(d\theta_B\we\theta^A).
\end{multline*}

The next bracket reads
\begin{multline}
\{B_2(a),S_2(M)\}=\int_\Sigma -\frac{1}{2}\eps^D{}_{BCA}\theta^B\we\theta^C\we (\xi^AdM+Md\xi^A) *(a\we p_D)+\\+M[(*\xi^B)\we\Star{A}(a\we p_B)]\we d\xi^A-\frac{1}{2}\eps^{D}{}_{BCA}{\theta}^{B}\we{\theta}^{C}\xi^A\we M a *dp_{D}
\label{b2-s2-0}
\end{multline}
---here we omitted two terms which are zero by virtue of \eqref{xi-d*=0}. To simplify the resulting expression let us first consider the two terms above containing $d\xi^A$---the first  of the terms can be transformed by means of \eqref{ett-al}:  
\[
-\frac{1}{2}\eps^D{}_{BCA}\theta^B\we\theta^C\we Md\xi^A*(a\we p_D)=M(*\xi^D)\vth_A\lr d\xi^A*(a\we p_D)=M(\vth_A\lr d\xi^A)a\we\xi^Dp_D.
\]
On the other hand by virtue of \eqref{abg-3} the other term  
\begin{equation*}
M[(*\xi^B)\we\Star{A}(a\we p_B)]\we d\xi^A=-M(*\xi^B)\we*(a\we p_B)(\vth_A\lr d\xi^A)=-M(\vth_A\lr d\xi^A)a\we\xi^Bp_B.
\end{equation*}
Thus the sum of the two terms in \eqref{b2-s2-0} containing $d\xi^A$ is zero. Consequently,
\begin{multline}
\{B_2(a),S_2(M)\}=\int_\Sigma -\frac{1}{2}\eps^D{}_{BCA}\theta^B\we\theta^C\xi^A\we(dM*(a\we p_D)+Ma*dp_D)=\\=\int_\Sigma*\theta^D\we (dM*(a\we p_D)+Ma*dp_D),
\label{b2-s2}
\end{multline}
 where we applied \eqref{ett-xi}.

Finally, the terms in \eqref{B-S} linear in $p_A$ read
\begin{multline}
\{B_1(a),S_1(M)\}+\{B_2(a),S_2(M)\}=\int_\Sigma*\theta^D\we (dM*(a\we p_D)+Ma*dp_D)+\\+Ma\we\theta^B\we*d\theta_A*(p_B\we\theta^A)+Md*(a\we\theta_A)\we\theta^B*(p_B\we\theta^A)-\\-Ma\we\theta^B\we*d\theta_B*(p_A\we\theta^A)-\frac{M}{2}d*(a\we\theta_A)\we\theta^A*(p_B\we\theta^B)+\\+M*(a\we\theta^B\we\theta^A)*p_A\we d\theta_B-Ma\we*p_A\we\theta^B*(d\theta_B\we\theta^A).
\label{BS-1}
\end{multline}

%***************************************************
\subsubsection{Terms independent of $p_A$ }
%***************************************************

The remaining three brackets $\{B_1(a),S_3(M)\}$, $\{B_1(a),S_2(M)\}$ and $\{B_2(a),S_3(M)\}$ give terms independent of $p_A$. The first bracket is zero since both $B_1(a)$ and $S_3(M)$ do not depend on $p_A$. The second bracket contains a term $d*(a\we\theta_A)\we d(M\xi^A)$ being an exact three-form---the integral of this term over $\Sigma$ is zero hence
\begin{multline*}
\{B_1(a),S_2(M)\}=\int_\Sigma [-a\we*d\theta_A+(a\we\theta^B)\we\Star{A}d\theta_B]\we(dM\xi^A+Md\xi^A)=\\=\int_\Sigma-dM\we a\we*(\xi^Ad\theta_A)-Ma\we*d\theta_A\we d\xi^A+Ma\we\theta^B\we d\xi^A*(d\theta_B\we\theta_A)-\\-M*(*d\theta_B\we\theta_A)\we*(a\we\theta^B)\we d\xi^A,
\end{multline*}
where in the second step we used the second line of \eqref{xi-d*=0}, \eqref{abg}, \eqref{a-*g} and \eqref{a-b}. The third bracket 
\begin{multline*}
\{B_2(a),S_3(M)\}=-\int_\Sigma d\Big(M\theta^B*(d\theta_B\we\theta_A)-\frac{M}{2}\theta_A*(d\theta_B\we\theta^B)\Big)\we a\xi^A+\\+M\Big(d\theta_B*(d\theta_A\we\theta^B)-\frac{1}{2}d\theta_A*(d\theta_B\we\theta^B)\Big)\we a\xi^A=-\int_\Sigma-Ma\we\theta^B\we\xi^Ad*(d\theta_B\we\theta_A)+\\+Ma\we \xi^Ad\theta_B*(d\theta_A\we\theta^B)-Ma\we\xi^A d\theta_A*(d\theta_B\we\theta^B),
\end{multline*}
---here in the first step we applied \eqref{xi-d*=0} and in the second one we carried out the exterior differentiation at the r.h.s. of the first line. Thus the terms in \eqref{B-S} independent of $p_A$ read
\begin{multline}
\{B_1(a),S_2(M)\}+\{B_2(a),S_3(M)\}=\int_\Sigma-dM\we a\we*(\xi^Ad\theta_A)-Ma\we*d\theta_A\we d\xi^A+\\+Ma\we\theta^B\we d\xi^A*(d\theta_B\we\theta_A)-M*(*d\theta_B\we\theta_A)\we*(a\we\theta^B)\we d\xi^A+\\+Ma\we\theta^B\we\xi^Ad*(d\theta_B\we\theta_A)-Ma\we \xi^Ad\theta_B*(d\theta_A\we\theta^B)+Ma\we\xi^A d\theta_A*(d\theta_B\we\theta^B).
\label{BS-0}
\end{multline}

%***************************************************
\subsubsection{Isolating constraints}
%***************************************************

Again, our goal now is to express the bracket 
\begin{multline*}
\{B(a),S(M)\}= \text{the terms \eqref{BS-2} quadratic in $p_A$}+\\+\text{the terms \eqref{BS-1} linear in $p_A$}+\text{the terms \eqref{BS-0} independent of $p_A$ }
\end{multline*}
as a sum of the constraints \eqref{B-sm}---\eqref{V-sm} smeared with some fields. 

\paragraph{Terms quadratic in $p_A$} We immediately see that the formula \eqref{BS-2} can be expressed as
\begin{multline}
-B\Big(M[\theta^B*(p_B\we a)-\frac{1}{2}a*(p_B\we\theta^B)]\Big)+\int_\Sigma M[\theta^B*(p_B\we a)-\frac{1}{2}a*(p_B\we\theta^B)]\we\theta^A\we*d\theta_A,
\label{BS-2a}
\end{multline}
hence
\begin{multline}
\{B(a),S(M)\}= -B\Big(M[\theta^B*(p_B\we a)-\frac{1}{2}a*(p_B\we\theta^B)]\Big)+\\+\text{terms linear in $p_A$}+\text{the terms \eqref{BS-0} independent of $p_A$},
\label{BS-B-1-0}
\end{multline}
where now the phrase ``terms linear in $p_A$'' means \eqref{BS-1} and the last term in \eqref{BS-2a}.

\paragraph{Terms linear in $p_A$} According to the last statement of the previous paragraph the remaining terms linear in $p_A$ read
\begin{multline}
\int_\Sigma M[\theta^B*(p_B\we a)-\frac{1}{2}a*(p_B\we\theta^B)]\we\theta^A\we*d\theta_A+*\theta^D\we (dM*(a\we p_D)+Ma*dp_D)+\\+Ma\we\theta^B\we*d\theta_A*(p_B\we\theta^A)+Md*(a\we\theta_A)\we\theta^B*(p_B\we\theta^A)-\\-Ma\we\theta^B\we*d\theta_B*(p_A\we\theta^A)-\frac{M}{2}d*(a\we\theta_A)\we\theta^A*(p_B\we\theta^B)+\\+M*(a\we\theta^B\we\theta^A)*p_A\we d\theta_B-Ma\we*p_A\we\theta^B*(d\theta_B\we\theta^A).
\label{BS-1-all}
\end{multline}

Let us now transform the terms containing $dM$ and $dp_D$ appearing in the first line of \eqref{BS-1-all}:   
\begin{multline*}
\int_\Sigma*\theta^D\we dM*(a\we p_D)=\int_\Sigma*(*dM\we\theta^D)a\we p_D\int_\Sigma=(\vth^D\lr dM)a\we p_D=\\=\int_\Sigma dM(\vth^D\lr a)\we p_D-dM\we a\vth^D\lr p_D=\int_\Sigma dM(\vth^D\lr a)\we p_D+*(dM\we a)\theta^A\we*p_A=\\=R\Big(*(dM\we a)\Big)+\int_\Sigma dM(\vth^D\lr a)\we p_D+*(dM\we a)\we \xi^Ad\theta_A,
\end{multline*}
where we  used \eqref{a-b}. On the other hand the term with $dp_D$ 
\begin{multline*}
\int_\Sigma *\theta^D\we Ma*dp_D =\int_\Sigma M*(*a\we\theta^D)dp_D=-\int_\Sigma d[M*(*a\we\theta^D)]\we p_D=\\=-\int_\Sigma dM(\vth^D\lr a)\we p_D+Md*(*a\we\theta^D)\we p_D.
\end{multline*}
Consequently, the sum of the two terms 
\begin{multline}
\int_\Sigma *\theta^D\we (dM*(a\we p_D)+Ma*dp_D)= R\Big(*(dM\we a)\Big)+\\+\int_\Sigma -Md*(*a\we\theta^A)\we p_A+dM\we a\we *(\xi^Ad\theta_A).
\label{dM-dp}
\end{multline}

The result just obtained means that \eqref{BS-B-1-0} can be re-expressed as
\begin{multline}
\{B(a),S(M)\}= -B\Big(M[\theta^B*(p_B\we a)-\frac{1}{2}a*(p_B\we\theta^B)]\Big)+R\Big(*(dM\we a)\Big)\\+\text{terms linear in $p_A$}+\text{terms independent of $p_A$},
\label{BS-BR-1-0}
\end{multline}
 where now $(i)$ the phrase ``terms linear in $p_A$'' means the second term at the r.h.s. of \eqref{dM-dp} and \eqref{BS-1-all} except the terms containing $dM$ and $dp_D$ and $(ii)$ ``terms independent of $p_A$'' means the last term in \eqref{dM-dp} and the terms \eqref{BS-0}.     

Note now that the form of constraints at the r.h.s. of \eqref{BS-BR-1-0} we managed to isolate so far resemble closely the form of the constraints at the r.h.s. of \eqref{RS-fin}. Let us then {\em assume} that
\begin{equation}
\{B(a),S(M)\}= -B\Big(M[\theta^B*(p_B\we a)-\frac{1}{2}a*(p_B\we\theta^B)+d\xi_B*(a\we*\theta^B)]\Big)+R\Big(*(dM\we a)\Big).
\label{BS-fin}
\end{equation} 
To justify the assumption we will proceed as follows: we will add to the r.h.s. of \eqref{BS-BR-1-0} zero expressed as
\begin{multline}
0=-B\Big(Md\xi_B*(a\we*\theta^B)\Big)+B\Big(Md\xi_B*(a\we*\theta^B)\Big)=-B\Big(Md\xi_B*(a\we*\theta^B)\Big)+\\+\int_\Sigma Md\xi_B*(a\we*\theta^B)\we(\theta^A\we*d\theta_A+\xi^Ap_A)=-B\Big(Md\xi_B*(a\we*\theta^B)\Big)+\\+\int_\Sigma Md\xi_B*(a\we*\theta^B)\theta^A\we*d\theta_A+Md*(*\theta_B\we\theta^A)\we p_A *(a\we*\theta^B)
\label{zero}
\end{multline}
(here in the last step we used \eqref{dtt-dxx}). Next we will show that all the remaining terms linear in $p_A$ sum up to zero, and that similarly do all the remaining terms independent of the momenta. Note that now the description of the terms linear in and independent of the momenta given just below Equation \eqref{BS-BR-1-0} has to be completed by taking into account the two last term in \eqref{zero}.

To demonstrate that all the remaining terms linear in $p_A$, that is, 
\begin{multline}
\int_\Sigma M\Big(\theta^B*(p_B\we a)\we\theta^A\we*d\theta_A-\frac{1}{2}a*(p_B\we\theta^B)\we\theta^A\we*d\theta_A-d*(*a\we\theta^A)\we p_A+\\+a\we\theta^B\we*d\theta_A*(p_B\we\theta^A)+d*(a\we\theta_A)\we\theta^B*(p_B\we\theta^A)-\\-a\we\theta^B\we*d\theta_B*(p_A\we\theta^A)-\frac{1}{2}d*(a\we\theta_A)\we\theta^A*(p_B\we\theta^B)+\\+*(a\we\theta^B\we\theta^A)*p_A\we d\theta_B-a\we*p_A\we\theta^B*(d\theta_B\we\theta^A)+d*(*\theta_B\we\theta^A)\we p_A *(a\we*\theta^B)\Big)
\label{BS-1-all-1}
\end{multline}
sum up to zero let us first perform some transformations. First we are going to show that the first, fourth, sixth and eighth terms above give together zero. To this end let us transform the eighth one as follows
\begin{multline*}
*(a\we\theta^B\we\theta^A)*p_A\we d\theta_B=-[\vth^B\lr*(a\we\theta^A)]*p_A\we d\theta_B=\\=-*(a\we\theta^A)*(p_A\we\theta^B)\we d\theta_B+*(a\we\theta^A)\we*p_A\we*(*d\theta_B\we\theta^B),
\end{multline*}    
where in the first step we applied \eqref{a-*g}, and in the second one we shifted the contraction $\vth^B\lr$ and used \eqref{a-*g} and \eqref{a-b}. Let us now transform the last term above in an analogous way:
\begin{multline*}
*(a\we\theta^A)\we*p_A\we*(*d\theta_B\we\theta^B)=\vth^A\lr *a\we *p_A\we*(*d\theta_B\we\theta^B)=\\=-*a\we*(p_A\we\theta^A)*(*d\theta_B\we\theta^B)+*a\we*p_A*(*d\theta_B\we\theta^B\we\theta^A)=\\=a\we\theta^B\we*d\theta_B*(p_A\we\theta^A)-\theta^A*(p_A\we a)\we\theta^B\we*d\theta_B.
\end{multline*}
Thus the eighth term
\begin{multline*}
*(a\we\theta^B\we\theta^A)*p_A\we d\theta_B=- a\we\theta^A\we *d\theta_B *(p_A\we\theta^B)+\\+a\we\theta^B\we *d\theta_B *(p_A\we\theta^A)-\theta^A*(p_A\we a)\we\theta^B\we*d\theta_B
\end{multline*}
and indeed the first, fourth, sixth and eighth terms disappear from \eqref{BS-1-all-1}.

Let us consider now the second and the seventh terms in \eqref{BS-1-all-1}. After a slight transformation of the second one their sum can be expressed as
\begin{multline*}
-\frac{1}{2}[*(a\we\theta^A)\we d\theta_A+d*(a\we\theta^A)\we\theta_A]*(p_B\we\theta^B)=\\=-\frac{1}{2}[2*(a\we\theta^A)\we d\theta_A-d(\theta_A*(a\we\theta^A))]*(p_B\we\theta^B)=\\-[*(a\we\theta^A)\we d\theta_A-d*a]*(p_B\we\theta^B),
\end{multline*}  
where in the last step we used \eqref{t*at}. 

Now the terms \eqref{BS-1-all-1} can be re-expressed in a simpler form as
\begin{multline*}
\int_\Sigma M\Big(-*(a\we\theta^A)\we d\theta_A*(p_B\we\theta^B)+(d*a)*(p_B\we\theta^B)-d*(*a\we\theta^A)\we p_A+\\+d*(a\we\theta_A)\we\theta^B*(p_B\we\theta^A)-a\we*p_A\we\theta^B*(d\theta_B\we\theta^A)+d*(*\theta_B\we\theta^A)\we p_A *(a\we*\theta^B)\Big).
\end{multline*}
Note that in each term above one can isolate the factor $p_B$ obtaining thereby 
\begin{multline}
\int_\Sigma Mp_B\we\Big(-\theta^B*[*(a\we\theta^A)\we d\theta_A] +\theta^B*d*a- d*(*a\we\theta^B)+\\+\theta^A*[d*(a\we\theta_A)\we\theta^B]+*(a\we\theta^A)*(d\theta_A\we\theta^B)+[d*(*\theta_A\we\theta^B)]*(a\we*\theta^A)\Big).
\label{BS-1-all-fin}
\end{multline}

Now using tensor calculus (see Section \ref{t-calc}) we will show that the terms in the big parenthesis above sum up to zero for every $\theta^A$ and $a$. The first term in \eqref{BS-1-all-fin} can be expressed as
\begin{multline}
-\theta^B*[*(a\we\theta^A)\we d\theta_A]=-\frac{1}{2}\theta^{B}_c(a\we\theta^A)^{ab}(d\theta_A)_{ab}dx^c=\\=-\theta^B_c a^a\theta^{Ab}(\nabla_a\theta_{Ab}-\nabla_b\theta_{Aa})dx^c= a^a\theta^{Ab}(\nabla_b\theta_{Aa})\theta^B_cdx^c, 
\label{BS-1-a}
\end{multline}
where in the first step we applied \eqref{*-df} and $*\eps=1$, and in the last one \eqref{nab-tt}.

The second term in \eqref{BS-1-all-fin} 
\begin{equation}
\theta^B*d*a=(\nabla^aa_a)\theta^B_c dx^c
\label{BS-1-b}
\end{equation}
by virtue of \eqref{delta-b}.

Due to \eqref{a-b} the third one 
\begin{equation}
- d*(*a\we\theta^B)=-\nabla_c(\theta^{Ba}a_a)dx^c=-(\nabla_c\theta^{Ba})a_adx^c-\theta^{Ba}(\nabla_c a_a)dx^c.
\label{BS-1-c}
\end{equation}

The fourth term in \eqref{BS-1-all-fin} by means of the last formula in \eqref{d-nabla} (set $\alpha=*(a\we\theta_A)$ and $\beta=\theta^B$), \eqref{eps-delta} and \eqref{q} can be expressed as
\begin{multline}
\theta^A*[d*(a\we\theta_A)\we\theta^B]=\nabla_d(a^a\theta^b_A)\theta^B_e\theta^A_c\eps_{abf}\eps^{dfe}dx^c=\\=-(\nabla_a a^a)\theta^B_c dx^c-a^a(\nabla_a\theta^b_A)\theta^B_b\theta^A_c dx^c+(\nabla_c a^a)\theta^B_a dx^c+a^a(\nabla_b\theta^b_A)\theta^B_a\theta^A_cdx^c.
\label{BS-1-d}
\end{multline}

The fifth one 
\begin{equation*}
*(a\we\theta^A)*(d\theta_A\we\theta^B)=a^a\theta^{Ab}(\nabla_d\theta_{Ae})\theta^B_f\eps_{abc}\eps^{def}dx^c.
\end{equation*}
Using \eqref{eps-delta} we obtain six terms and two of them vanish by virtue of \eqref{nab-tt}. Thus
\begin{multline}
*(a\we\theta^A)*(d\theta_A\we\theta^B)=a^a\theta^{Ab}(\nabla_b\theta_{Ac})\theta^B_a dx^c+a^a\theta^{Ab}(\nabla_c\theta_{Aa})\theta^B_b dx^c-\\-a^a\theta^{Ab}(\nabla_b\theta_{Aa})\theta^B_cdx^c -a^a\theta^{Ab}(\nabla_a\theta_{Ac})\theta^B_b dx^c.
\label{BS-1-e}
\end{multline}

Due to \eqref{a-b} and \eqref{q} the last term in \eqref{BS-1-all-fin} 
\begin{multline}
[d*(*\theta_A\we\theta^B)]*(a\we*\theta^A)=\nabla_c(\theta^b_A\theta^B_b)a_a\theta^{Aa}dx^c=(\nabla_c\theta^b_A)\theta^B_b a_a\theta^{Aa}dx^c+\\+(\nabla_c \theta^{Ba})a_adx^c.
\label{BS-1-f}
\end{multline}

Collecting all the results \eqref{BS-1-a}--\eqref{BS-1-f} we note that we obtain two kinds of terms: ones containing a covariant derivative of $a^a$ and ones containing a covariant derivative of $\theta^A_a$. The terms containing $\nabla_b a^a$ appear in \eqref{BS-1-b}, \eqref{BS-1-c} and \eqref{BS-1-d} and sum up to zero:  
\begin{equation*}
(\nabla^aa_a)\theta^B_c dx^c-\theta^{Ba}(\nabla_c a_a)dx^c-(\nabla_a a^a)\theta^B_c dx^c+(\nabla_c a^a)\theta^B_a dx^c=0.
\end{equation*}
Regarding the terms containing  $\nabla_a\theta^A_b$, there are ten of them and they can be grouped into pairs such that each pair sums up to zero. These pairs are:

\begin{enumerate}
\item the term at the r.h.s. of \eqref{BS-1-a} and the third term at the r.h.s. of \eqref{BS-1-e}, 
\item the first term at the r.h.s. of \eqref{BS-1-c} and the last term at the r.h.s. of \eqref{BS-1-f}, 
\item the second term at the r.h.s. of \eqref{BS-1-d} and the last term at the r.h.s. of \eqref{BS-1-e} (shift the derivative by means of \eqref{0-nab-q} in the latter term), 
\item the last term at the r.h.s. of \eqref{BS-1-d} and the first term at the r.h.s. of \eqref{BS-1-e} (shift the derivative by means of \eqref{0-nab-q} in the latter term), 
\item the second term at the r.h.s. of \eqref{BS-1-e} and the first term at the r.h.s. of \eqref{BS-1-f} (again shift the derivative by means of \eqref{0-nab-q} in the latter term). 
\end{enumerate}

\paragraph{Terms independent of $p_A$} Our goal now is to show that all the remaining terms independent of the momenta i.e. the terms \eqref{BS-0}, the last term in \eqref{dM-dp} and the second term at the r.h.s. of \eqref{zero} sum up to zero. Note that the first term in \eqref{BS-0} cancels the last term  in \eqref{dM-dp}. Now in all remaining terms there is the factor $Ma$ and therefore they can be expressed as 
\begin{multline}
\int_\Sigma Ma\we\Big( -*d\theta_A\we d\xi^A+\theta^B*(d\theta_B\we\theta_A)\we d\xi^A+\theta^B\we*[*(*d\theta_B\we\theta_A)\we d\xi^A]+\\+\theta^B\we\xi^Ad*(d\theta_B\we\theta_A)-\xi^Ad\theta_B*(d\theta_A\we\theta^B)+\xi^Ad\theta_A*(d\theta_B\we\theta^B)+\\+*\theta^A*(d\xi_A\we\theta^B\we*d\theta_B)\Big)
\label{ind-pa}
\end{multline}
In the fourth, fifth and sixth terms above there appears the function $\xi^A$ while in the remaining ones there is the derivative $d\xi^A$. Let us then transform the three terms to obtain ones containing $d\xi^A$. To transform the fourth one we note that
\begin{multline*}
0=-d\big(\theta^B\we\xi^A*(d\theta_B\we\theta_A)\big)=-d\theta^B\we\xi^A*(d\theta_B\we\theta_A)+\theta^B\we d\xi^A*(d\theta_B\we\theta_A)+\\+\theta^B\we\xi^Ad*(d\theta_B\we\theta_A)=\theta^B*(d\theta_B\we\theta_A)\we d\xi^A+\theta^B\we\xi^Ad*(d\theta_B\we\theta_A),
\end{multline*}   
which means that the sum of the second and the fourth term in \eqref{ind-pa} is zero. The fifth term in \eqref{ind-pa} 
\[
-\xi^Ad\theta_B*(d\theta_A\we\theta^B)=-d\theta_B*(\xi^Ad\theta_A\we\theta^B)=-d\theta_B*(\theta_A\we d\xi^A\we\theta^B).
\]
Transforming similarly the sixth term we can rewrite \eqref{ind-pa} as follows:
\begin{multline*}
\int_\Sigma Ma\we\Big( -*d\theta_A\we d\xi^A+\theta^B\we*[*(*d\theta_B\we\theta_A)\we d\xi^A]-d\theta_B*(\theta_A\we d\xi^A\we\theta^B)+\\+\theta_A\we d\xi^A*(d\theta_B\we\theta^B)+*\theta^A*(d\xi_A\we\theta^B\we*d\theta_B)\Big)
\end{multline*}
By a direct calculation using tensor calculus we will demonstrate that the terms in the big parenthesis sum up to zero for all $\theta^A$. More precisely, we will show that 
\begin{multline}
*\Big( -*d\theta_A\we d\xi^A+\theta^B\we*[*(*d\theta_B\we\theta_A)\we d\xi^A]-d\theta_B*(\theta_A\we d\xi^A\we\theta^B)+\\+\theta_A\we d\xi^A*(d\theta_B\we\theta^B)+*\theta^A*(d\xi_A\we\theta^B\we*d\theta_B)\Big)
\label{BS-0-fin}
\end{multline}
is equal to zero.

The first term in the expression above  reads by virtue of \eqref{a-b}
\begin{equation}
-*(*d\theta_A\we d\xi^A)=-\overrightarrow{d\xi^A}\lr d\theta_A=-(\nabla^a\xi^A)(\nabla_a\theta_{Ab})dx^b+(\nabla^a\xi^A)(\nabla_b\theta_{Aa})dx^b.
\label{BS-0-a}
\end{equation}

Using twice \eqref{a-b} we express the  second term in \eqref{BS-0-fin} as follows
\begin{multline}
*\big(\theta^B\we*[*(*d\theta_B\we\theta_A)\we d\xi^A]\big)=-\vth^B\lr[\vth_A\lr d\theta_B\we d\xi^A]=\theta^a_A\theta^{Bb}(\nabla_b\theta_{Ba})(\nabla_c\xi^A)dx^c+\\+\theta^a_A(\nabla_a\theta_{Bb})\theta^{Bc}(\nabla_c\xi^A)dx^b-\theta^a_A(\nabla_b\theta_{Ba})\theta^{Bc}(\nabla_c\xi^A)dx^b,
\label{BS-0-c}
\end{multline}
where in the second step we applied \eqref{nab-tt}.

The third term 
\[
-*d\theta_B*(\theta_A\we d\xi^A\we\theta^B)=-\theta_{Aa}(\nabla_b\xi^A)\theta^B_c(\nabla^d\theta^e_B)\eps^{abc}\eps_{def}dx^f.
\]
Applying \eqref{eps-delta} we again obtain six terms, two of them vanish by virtue of \eqref{nab-tt} and we are left with the following expression
\begin{multline}
-*d\theta_B*(\theta_A\we d\xi^A\we\theta^B)=-\theta_{Aa}(\nabla_b\xi^A)\theta^B_c(\nabla^a\theta^b_B)dx^c-\theta_{Aa}(\nabla_b\xi^A)\theta^B_c(\nabla^c\theta^a_B)dx^b+\\+\theta_{Aa}(\nabla_b\xi^A)\theta^B_c(\nabla^b\theta^a_B)dx^c+\theta_{Aa}(\nabla_b\xi^A)\theta^B_c(\nabla^c\theta^b_B)dx^a.
\label{BS-0-e}
\end{multline}

The fourth term in \eqref{BS-0-fin} reads
\begin{multline}
*(\theta_A\we d\xi^A)*(d\theta_B\we\theta^B)=\theta_A^a(\nabla^b\xi^A)(\nabla_d\theta_{Be})\theta^B_f\eps^{def}\eps_{abc} dx^c=\theta_A^a(\nabla^b\xi^A)(\nabla_a\theta_{Bb})\theta^B_c dx^c+\\+\theta_A^a(\nabla^b\xi^A)(\nabla_b\theta_{Bc})\theta^B_a dx^c+\theta_A^a(\nabla^b\xi^A)(\nabla_c\theta_{Ba})\theta^B_b dx^c-\theta_A^a(\nabla^b\xi^A)(\nabla_b\theta_{Ba})\theta^B_c dx^c-\\-\theta_A^a(\nabla^b\xi^A)(\nabla_a\theta_{Bc})\theta^B_b dx^c-\theta_A^a(\nabla^b\xi^A)(\nabla_c\theta_{Bb})\theta^B_a dx^c.
\label{BS-0-f}
\end{multline}

Finally, the last term in \eqref{BS-0-fin} 
\begin{equation}
\theta^A*(d\xi_A\we\theta^B\we*d\theta_B)=(\nabla_d\xi_A)\theta^B_e(\nabla^a\theta^b_B)\theta^A_c\eps_{abf}\eps^{def}dx^c=-(\nabla_b\xi_A)\theta^B_a(\nabla^a\theta^b_B)\theta^A_cdx^c,
\label{BS-0-p}
\end{equation}
where in the last step we used \eqref{eps-delta} and \eqref{nab-tt}.

In this way we managed to express \eqref{BS-0-fin} in terms of the components $\theta^A_a$ and $\xi^A$ and their covariant derivatives obtaining altogether sixteen terms \eqref{BS-0-a}--\eqref{BS-0-p}. As before those terms can be grouped into pairs such that terms in each pair sum up to zero. Let us now enumerate the pairs:

\begin{enumerate}
\item the first term at the r.h.s. of \eqref{BS-0-a} and the second term at the r.h.s. of \eqref{BS-0-f}. Here the latter term needs the following transformation:
\begin{multline*}
\theta_A^a(\nabla^b\xi^A)(\nabla_b\theta_{Bc})\theta^B_a dx^c=(\nabla^b\xi^A)(\nabla_b\theta_{Bc})(\delta^B{}_A+\xi^B\xi_A)dx^c=\\=(\nabla^b\xi^A)(\nabla_b\theta_{Ac})dx^c,
\end{multline*}
where in the first step we used \eqref{tt-eta}, the second one holds true by virtue of $(\nabla^b\xi^A)\xi_A=0$.  
\item the second term at the r.h.s. of \eqref{BS-0-a} and the sixth term at the r.h.s. of \eqref{BS-0-f} (the latter term needs a transformation analogous to that shown above),
\item the first term at the r.h.s. of \eqref{BS-0-c} and the second term at the r.h.s. of \eqref{BS-0-e},
\item the second term at the r.h.s. of \eqref{BS-0-c} and the first term at the r.h.s. of \eqref{BS-0-f} (apply \eqref{0-nab-q} to the latter term),
\item the third term at the r.h.s. of \eqref{BS-0-c} and the third term at the r.h.s. of \eqref{BS-0-f},
\item the first term at the r.h.s. of \eqref{BS-0-e} and the fifth term at the r.h.s. of \eqref{BS-0-f} (apply \eqref{0-nab-q} to the latter term),
\item the third term at the r.h.s. of \eqref{BS-0-e} and the fourth term at the r.h.s. of \eqref{BS-0-f},
\item the fourth term at the r.h.s. of \eqref{BS-0-e} and the term at the r.h.s. of \eqref{BS-0-p}.
\end{enumerate}  

Thus we managed to demonstrate that all the remaining terms \eqref{BS-0-fin} independent of $p_A$ sum up to zero and thereby proved the assumption \eqref{BS-fin}.

%***************************************************
\subsection{Poisson brackets of $V(\vec{M})$}
%***************************************************

The functional derivatives of the smeared scalar constraint $V(\vec{M})$ (see \eqref{V-sm}) are of the following form \cite{os}: 
\begin{align}
\frac{\delta V(\vec{M})}{\delta {\theta}^A}&=-{\cal L}_{\vec{M}}p_A, & \frac{\delta V(\vec{M})}{\delta p_A}&={\cal L}_{\vec{M}}{\theta}^A.
\label{delta-V}
\end{align}
It was shown in \cite{os} that
\begin{equation}
\{V(\vec{M}),V(\vec{M}')\}=V([\vec{M},\vec{M}'])=V(\LL_{\vec{M}}\vec{M}'),
\label{VV-fin}
\end{equation}
where $[\vec{M},\vec{M}']$ denotes the Lie bracket of the vector fields $\vec{M},\vec{M}'$ on $\Sigma$.  

Derivations of brackets of $V(\vec{M})$ and the other constraints will be based on the following formula \cite{os}:
\begin{equation}
\LL_{\vec{M}}(\alpha\we*\beta)=\LL_{\vec{M}}\alpha\we*\beta+\alpha\we*\LL_{\vec{M}}\beta+\LL_{\vec{M}}\theta^A\we(\alpha\we\Star{A}\beta).
\label{L-*}
\end{equation}
We will also apply the following well known properties of the Lie derivative:
\begin{align}
0&=\int_{\Sigma}\LL_{\vec{M}}(\alpha\we\beta)=\int_\Sigma(\LL_{\vec{M}}\alpha)\we\beta+\int_\Sigma\alpha\we\LL_{\vec{M}}\beta,\label{LM-a}\\ 
d(\LL_{\vec{M}}\gamma)&=\LL_{\vec{M}}(d\gamma)\label{dLLd}, 
\end{align}
where $\alpha\we\beta$ is a three-form, and $\gamma$ any $k$-form on $\Sigma$.     

%***************************************************
\subsubsection{Poisson bracket of $V(\vec{M})$ and $S(M)$ }
%***************************************************

To calculate the bracket $\{V(\vec{M}),S(M)\}$ we will use the split of $S(M)$ into the three functionals \eqref{S_i}:
\[
\{V(\vec{M}),S(M)\}=\sum_{i=1}^3\{V(\vec{M}),S_i(M)\}.
\]   

To calculate the first term $\{V(\vec{M}),S_1(M)\}$ at the r.h.s. of this equation let us split $S_1(M)$ into a sum $S_1(M)=S_{11}(M)+S_{12}(M)$, where 
\begin{align*}
S_{11}(M)&:=\int_\Sigma \frac{M}{2}(p_A\we\theta^B)\we*(p_B\we\theta^A), & S_{12}(M)&:=-\int_\Sigma \frac{M}{4}(p_A\we\theta^A)\we*(p_B\we\theta^B)
\end{align*}
and consider the bracket $\{V(\vec{M}),S_{11}(M)\}$: 
\begin{multline*}
\{V(\vec{M}),S_{11}(M)\}=\int_\Sigma -{\cal L}_{\vec{M}}p_A\we M\theta^B*(p_B\we\theta^A)-\\-M\Big(p_B*(p_A\we\theta^B)+\frac{1}{2}(p_C\we\theta^B)\we\Star{A}(p_B\we\theta^C)\Big)\we {\cal L}_{\vec{M}}{\theta}^A=\\=-\int_\Sigma M\Big(\LL_{\vec{M}}(p_A\we\theta^B)\we*(p_B\we\theta^A)+{\cal L}_{\vec{M}}{\theta}^A\we\frac{1}{2}[(p_C\we\theta^B)\we\Star{A}(p_B\we\theta^C)]\Big)
\end{multline*}
---functional derivatives of $S_{11}(M)$ used to calculate the bracket can be read off from \eqref{S1/t} and \eqref{S1/p}. It is easy to see that the first term in the last line above
\[
M\LL_{\vec{M}}(p_A\we\theta^B)\we*(p_B\we\theta^A)=\frac{M}{2}\Big(\LL_{\vec{M}}(p_A\we\theta^B)\we*(p_B\we\theta^A)+(p_A\we\theta^B)\we*\LL_{\vec{M}}(p_B\we\theta^A)\Big).
\]
This fact together with \eqref{L-*} allow us to write
\begin{multline}
\{V(\vec{M}),S_{11}(M)\}=-\int_\Sigma M\LL_{\vec{M}}\Big(\frac{1}{2}(p_A\we\theta^B)\we*(p_B\we\theta^A)\Big)=\\=\int_\Sigma (\LL_{\vec{M}}M)\Big(\frac{1}{2}(p_A\we\theta^B)\we*(p_B\we\theta^A)\Big)=S_{11}(\LL_{\vec{M}}M),
\label{VS11}
\end{multline}
where the second equality holds by virtue of \eqref{LM-a}. It can be shown in a similar way that 
\begin{equation}
\{V(\vec{M}),S_{12}(M)\}=S_{12}(\LL_{\vec{M}}M).
\label{VS12}
\end{equation}

According to calculations carried out in \cite{os} 
\begin{equation}
\{V(\vec{M}),S_{2}(M)\}=S_{2}(\LL_{\vec{M}}M).
\label{VS2}
\end{equation}

Let us split $S_3(M)$ as follows:  $S_3(M)=S_{31}(M)+S_{32}(M)$, where
\begin{align*}
S_{31}(M)&:=\int_\Sigma \frac{M}{2}(d\theta_A\we\theta^B)\we{*}(d\theta_B\we\theta^A),\\S_{32}(M)&:=-\int_\Sigma \frac{M}{4}(d\theta_A\we\theta^A)\we{*}(d\theta_B\we\theta^B).
\end{align*}
Reading off from \eqref{S3/t} and \eqref{S3/p} functional derivatives of $S_{31}(M)$ we obtain
\begin{multline}
\{V(\vec{M}),S_{31}(M)\}=-\int_\Sigma d\big(M\theta^B*(d\theta_B\we\theta_A)\big)\we\LL_{\vec{M}}\theta^A+M d\theta_B*(d\theta_A\we\theta^B)\we\LL_{\vec{M}}\theta^A\\+\frac{1}{2}\big((d\theta_C\we\theta^B)\we\Star{A}(d\theta_B\we\theta^C)\Big)\we\LL_{\vec{M}}\theta^A 
\label{VS31-0}
\end{multline}
The first term at the r.h.s. of the formula above 
\[
\int_\Sigma d\big(M\theta^B*(d\theta_B\we\theta_A)\big)\we\LL_{\vec{M}}\theta^A=\int_\Sigma M\theta^B*(d\theta_B\we\theta_A)\we \LL_{\vec{M}}d\theta^A,
\]
where we shifted the derivative $d$ and applied \eqref{dLLd}. Thus the sum of the first two terms  at the r.h.s. of \eqref{VS31-0} reads
\begin{multline*}
\int_\Sigma M\LL_{\vec{M}}(d\theta^A\we\theta^B)*(d\theta_B\we\theta_A)=\\=\int_\Sigma \frac{M}{2}\big(\LL_{\vec{M}}(d\theta^A\we\theta^B)\we *(d\theta_B\we\theta_A)+d\theta^A\we\theta^B\we *\LL_{\vec{M}}(d\theta_B\we\theta_A)\big).
\end{multline*}
Setting this result to \eqref{VS31-0}, applying \eqref{L-*} and \eqref{LM-a} we obtain
\begin{equation}
\{V(\vec{M}),S_{31}(M)\}=\int_\Sigma (\LL_{\vec{M}}M)d\theta^A\we\theta^B\we *(d\theta_B\we\theta_A)=S_{31}({\LL_{\vec{M}}M}).
\label{VS31}
\end{equation}
In analogous way one can show that
\begin{equation}
\{V(\vec{M}),S_{32}(M)\}=S_{32}(\LL_{\vec{M}}M)
\label{VS32}
\end{equation}

Gathering all partial results \eqref{VS11}-\eqref{VS32} (except \eqref{VS31-0}) we obtain
\begin{equation}
\{V(\vec{M}),S(M)\}=S(\LL_{\vec{M}}M).
\label{VS-fin}
\end{equation}

%***************************************************
\subsubsection{Poisson bracket of $V(\vec{M})$ and the constraints $B(a)$ and $R(b)$}
%***************************************************

Here we explicitely calculate the bracket $\{V(\vec{M}),B(a)\}$. The bracket $\{V(\vec{M}),R(b)\}$ can be calculated similarly. 

Obviously,
\[
\{V(\vec{M}),B(a)\}=\{V(\vec{M}),B_1(a)\}+\{V(\vec{M}),B_2(a)\},
\]  
where $B_1(a)$ and $B_2(a)$ are given by \eqref{B1B2}. We have
\begin{multline}
\{V(\vec{M}),B_1(a)\}=-\int_\Sigma- a\we*d\theta_A\we{\cal L}_{\vec{M}}{\theta}^A+[(a\we\theta^B)\we\Star{A}d\theta_B]\we{\cal L}_{\vec{M}}{\theta}^A+\\+d*(a\we\theta_A)\we{\cal L}_{\vec{M}}{\theta}^A.
\label{VB1-0}
\end{multline}
The first term at the r.h.s. above
\[
\int_\Sigma-a\we*d\theta_A\we{\cal L}_{\vec{M}}{\theta}^A=\int_\Sigma {\cal L}_{\vec{M}}(a\we{\theta}^A)\we*d\theta_A-{\cal L}_{\vec{M}}a\we{\theta}^A\we*d\theta_A
\]
and the last term in \eqref{VB1-0}
\begin{multline*}
\int_\Sigma d*(a\we\theta_A)\we {\cal L}_{\vec{M}}{\theta}^A=\int_\Sigma  *(a\we\theta_A)\we d{\cal L}_{\vec{M}}{\theta}^A=\int_\Sigma *(a\we\theta_A)\we {\cal L}_{\vec{M}}d{\theta}^A =\\=\int_\Sigma a\we\theta_A\we *{\cal L}_{\vec{M}}d{\theta}^A
\end{multline*}
---here in the second step we used \eqref{dLLd}. Setting these two results to \eqref{VB1-0} and applying \eqref{L-*} and \eqref{LM-a} we obtain
\begin{multline}
\{V(\vec{M}),B_1(a)\}=-\int_{\Sigma} -\LL_{\vec{M}}a\we\theta^A\we*d\theta_A+\LL_{\vec{M}}(a\we\theta^A\we*d\theta_A)=\\=\int_\Sigma \LL_{\vec{M}}a\we \theta^A\we*d\theta_A=B_1(\LL_{\vec{M}}a). 
\label{VB1}
\end{multline}

The other bracket
\begin{multline}
\{V(\vec{M}),B_2(a)\}=\int_\Sigma -{\cal L}_{\vec{M}}p_A\we a\xi^A+ \frac{1}{2}\eps^D{}_{BCA}\theta^B\we\theta^C*(a\we p_D)\we{\cal L}_{\vec{M}}{\theta}^A-\\-[(*\xi^B)\we\Star{A}(a\we p_B)]\we{\cal L}_{\vec{M}}{\theta}^A.
\label{VB2-0}
\end{multline}
The first term at the r.h.s. above
\begin{multline*}
-{\cal L}_{\vec{M}}p_A\we a\xi^A=\xi^A\LL_{\vec{M}}a\we p_A-\xi^A\LL_{\vec{M}}(a\we p_A)=\LL_{\vec{M}}a\we\xi^A p_A-(*\xi^A)\we*\LL_{\vec{M}}(a\we p_A)
\end{multline*}
and the second one at the r.h.s. of \eqref{VB2-0}
\begin{multline*}
\frac{1}{2}\eps^D{}_{BCA}\theta^B\we\theta^C\we{\cal L}_{\vec{M}}{\theta}^A*(a\we p_D)=\frac{1}{3}\LL_{\vec{M}}\Big(\frac{1}{2}\eps^D{}_{BCA}\theta^B\we\theta^C\we\theta^A\Big)*(a\we p_D)=\\=-\LL_{\vec{M}}(*\xi^D)\we*(a\we p_D),
\end{multline*}
where in the last step we used \eqref{xi-sol}. Setting these two results to \eqref{VB1-0} and applying \eqref{L-*} and \eqref{LM-a} we obtain
\begin{equation*}
\{V(\vec{M}),B_2(a)\}=\int_\Sigma \LL_{\vec{M}}a\we\xi^A p_A-\LL_{\vec{M}}[*\xi^A\we*(a\we p_A)]=\int_\Sigma \LL_{\vec{M}}a\we\xi^A p_A=B_2(\LL_{\vec{M}}a).
\end{equation*}

The equation above and \eqref{VB1} give us the final result
\begin{equation}
\{V(\vec{M}),B(a)\}=B(\LL_{\vec{M}}a).
\label{VB-fin}
\end{equation}

Similarly,
\begin{equation}
\{V(\vec{M}),R(b)\}=R(\LL_{\vec{M}}b).
\label{VR-fin}
\end{equation}

%***************************************************
\section{Summary}
%***************************************************

To summarize the calculation let us list the results. The Poisson brackets of boosts and rotation constrains (Equations \eqref{BB-fin}, \eqref{RR-fin} and \eqref{BR-fin}):
\[
\begin{aligned}
&\{B(a),B(a')\}=-R(*(a\we a')),\\
&\{R(b),R(b')\}=R(*(b\we b')),\\
&\{B(a),R(b)\}=B(*(a\we b)).
\end{aligned}
\]
The bracket of the scalar constraints (Equation \eqref{SS-fin}):
\begin{multline*}
\{S(M),S(M')\}=V(\vec{m})+B\Big(\theta^B*(m\we p_B)-\frac{1}{2}*(m\we\xi^B*d\theta_B)-\\-*[m\we*(\theta^B\we*p_B)]-\frac{1}{2}*(*m\we\theta^B)*p_B+\frac{1}{2}*[*(m\we\theta^B)\we*p_B]\Big)+\\+R\Big(-\theta^B*(m\we d\theta_B)-\frac{1}{2}*(m\we\xi^B*p_B)+\\+*[m\we*(\theta^B\we*d\theta_B)]+\frac{1}{2}*(*m\we\theta^B)*d\theta_B-\frac{1}{2}*[*(m\we\theta^B)\we*d\theta_B]\Big).
\end{multline*}
In this formula
\[
m:=MdM'-M'dM.
\]
The brackets of  the boost and rotation constraints and the scalar one (Equations \eqref{BS-fin} and \eqref{RS-fin}): 
\begin{align*}
\{B(a),S(M)\}=&-B\Big(M[\theta^B*(p_B\we a)-\frac{1}{2}a*(p_B\we\theta^B)+d\xi_B*(a\we*\theta^B)]\Big)+\nonumber\\&+R\Big(*(dM\we a)\Big),\\ 
\{R(b),S(M)\}=&-R\Big(M[\theta^B*(p_B\we b)-\frac{1}{2}b*(p_B\we\theta^B)+d\xi_A*(b\we*\theta^A)]\Big)-\nonumber\\&-B\Big(*(dM\we b)\Big). 
\end{align*}
The brackets of the vector constraint (Equations \eqref{VV-fin}, \eqref{VS-fin}, \eqref{VB-fin} and \eqref{VR-fin}):
\[
\begin{aligned}
\{V(\vec{M}),V(\vec{M}')\}=&V(\LL_{\vec{M}}\vec{M}')\equiv V([\vec{M},\vec{M}']),\\
\{V(\vec{M}),S(M)\}=&S(\LL_{\vec{M}}M),\\
\{V(\vec{M}),B(a)\}=&B(\LL_{\vec{M}}a),\\
\{V(\vec{M}),R(b)\}=&R(\LL_{\vec{M}}b).
\end{aligned} 
\]
In the formulae above $\LL_{\vec{M}}$ denotes the Lie derivative on $\Sigma$ with respect to the vector field $\vec{M}$.  

A discussion of the results can be found in \cite{oko-tegr-I}, here we restrict ourselves to a statement that a Poisson bracket of every pair of the constraints \eqref{B-sm}---\eqref{V-sm} is a sum of the constraints smeared with some fields. In other words, the constraint algebra presented above is {\em closed}.

\paragraph{Acknowledgments} I am grateful to J\k{e}drzej \'Swie\.zewski for his cooperation in the research on a Hamiltonian model described in \cite{os} which was for me a preparatory exercise for deriving the results described in this paper. I am also grateful to Jerzy Lewandowski for a valuable discussion.

%***************************************************
%***************************************************

%***************************************************


\begin{thebibliography}{}
%***************************************************
%***************************************************


%
\bibitem{oko-tegr-I} Oko{\l}\'ow A 2011 ADM-like Hamiltonian formulation of gravity in the teleparallel geometry {\em accepted for publication in Gen. Rel. Grav.} ({\it E-print} \verb+arXiv:1111.5498v2+)
%
\bibitem{q-suit} Oko{\l}\'ow A 2013 Variables suitable for constructing quantum states for the Teleparallel Equivalent of General Relativity I {\it E-print} \verb+arXiv:1305.4526+ 
%
\bibitem{ham-nv} Oko{\l}\'ow A, Variables suitable for constructing quantum states for the Teleparallel Equivalent of General Relativity II {\em E-print} \verb+arXiv:1308.2104+
%
\bibitem{q-tegr} Oko{\l}\'ow A 2013 Kinematic quantum states for the Teleparallel Equivalent of General Relativity {\it E-print} \verb+arXiv:1304.6492+
%
\bibitem{maluf-1} Maluf J W, da Rocha-Neto J F 2001 Hamiltonian formulation of general relativity in the teleparallel geometry {\it Phys. Rev. D} {\bf 64} 084014 {\it E-print} \verb+arXiv:gr-qc/0002059+
%
\bibitem{maluf} da Rocha-Neto J F, Maluf J W, Ulhoa S C 2010 Hamiltonian formulation of unimodular gravity in the teleparallel geometry {\it Phys. Rev. D} {\bf 82} 124035 {\it E-print} \verb+arXiv:1101.2425+
%
\bibitem{bl} Blagojevi\'c M, Nikoli\'c I A 2000 Hamiltonian structure of the teleparallel formulation of GR {\it Phys. Rev. D} {\bf 62} 024021 {\it E-print} \verb+arXiv:hep-th/0002022+
% 
\bibitem{mielke} Mielke E W 1992 Ashtekar's Complex Variables in General Relativity and Its Teleparallelism Equivalent {\it Ann. Phys.} {\bf 219} 78-108
%
\bibitem{os} Oko{\l}\'ow A, \'Swie\.zewski J 2012 Hamiltonian formulation of a simple theory of the teleparallel geometry {\it Class. Quant. Grav.} {\bf 29} 045008 {\it E-print} \verb+arXiv:1111.5490+ 
%
\bibitem{nester} Nester J M 1989 Positive energy via the teleparallel Hamiltonian {\it Int. J. Mod. Phys. A} {\bf 4} 1755-1772



%***************************************************
\end{thebibliography}
\end{document}